\documentclass[a4paper]{JHEP3} %  

\usepackage{bbm,amsmath,graphicx,amssymb}
\usepackage{bm}
\usepackage{cite}
\usepackage{verbatim}
\usepackage{slashed}
\usepackage{epsf}

\def\spa#1.#2{\left\langle#1\,#2\right\rangle}
\def\spb#1.#2{\left[#1\,#2\right]}

\def\nn{\nonumber}

\def\cA{\mathcal{A}}

\def\stamp{--- {\bf \today} --- {\bf \jobname.tex}}

\def\fs_#1{\mathfrak{s}(#1)}

\def\BE{\begin{equation}}
\def\EE{\end{equation}}
\def\spa#1.#2{\left\langle#1\,#2\right\rangle}
\def\spb#1.#2{\left[#1\,#2\right]}
\def\lor#1.#2{\left(#1\,#2\right)}

\newcommand\fverb{\setbox\fverbbox=\hbox\bgroup\verb}
\newcommand\fverbdo{\egroup\medskip\noindent%
			\fbox{\unhbox\fverbbox}\ }
\newcommand\fverbit{\egroup\item[\fbox{\unhbox\fverbbox}]}
\newbox\fverbbox

\title{Perturbative Gravity and Gauge Theory Relations \\ -- A Review}

\author{Thomas S{\o}ndergaard\bigskip\\
{\small Kavli Institute for Theoretical Physics,\\
University of California, \\ Santa Barbara CA 93106,\\
USA \\ \\
\small Niels Bohr International Academy and Discovery Center,\\
The Niels Bohr Institute, \\ Blegdamsvej 17, DK-2100 Copenhagen,\\
Denmark\footnote{Address after July 9th, 2011.} \\ \\
{\tt email:}
tsonderg@nbi.dk
}}

\received{\today} 		%%
\accepted{\today}		%% These are for published papers.

\abstract{This review is dedicated to the amazing Kawai-Lewellen-Tye relations, connecting perturbative
gravity and gauge theories at tree level. The main focus is on $n$-point derivations and general properties
both from a string theory and pure field theory point of view. In particular the field theory part is based on some
very recent developments.
}


\begin{document}

\section{Introduction}
In 1985 Kawai, Lewellen and Tye (KLT) derived an amazing relation between gravity and gauge theory tree-level amplitudes
\cite{KLT,KLT2,Bern:1998sv,Bern:1999bx,Abe:2005se,KLTproofShort,KLTproofLong,Feng:2010br,Skernel,Du:2011js}. This was done
by factorizing a closed string into a sum of products between two open strings. As such it was a relation satisfied to all orders in $\alpha'$,
even when taking the field theory limit $\alpha' \rightarrow 0$ \cite{Green:1982sw,Berends:1988zp}. In particular, the validity in this limit has been a major puzzle for many years.
At the Lagrangian level any connection between
Einstein gravity and Yang-Mills theory seems highly unlikely. Expanding the Einstein-Hilbert Lagrangian perturbatively leads to an infinite series of
more and more complicated interaction terms, while Yang-Mills only involve three- and four-point interactions. Nevertheless, we will in this review see
how the KLT relations can be understood from a field theoretical point of view.

The new light that has recently been shed on these relations
is only a small part of a remarkable progress that is currently happening in
our understanding of scattering amplitudes, see \textit{e.g.} \cite{Cachazo:2005ga,Bern:2007dw,Dixon:2011xs} for reviews on some of these developments.
Indeed, as we will see, our new knowledge about KLT is to a large extent directly built
upon several of the great discoveries that have been made within recent years.

Inspired by Witten's famous 2004 paper \cite{Witten2004}, Britto,
Cachazo and Feng (BCF) uncovered an on-shell recursion relation for tree amplitudes from which one could construct higher-point amplitudes
from lower-point amplitudes \cite{BCF}. Together with Witten they soon after proved these 
\textit{BCFW recursion relations} \cite{BCFW}, which will play an important role for us. By now such recursion relations have also been
extended to string theory \cite{Cheung:2010vn} and even to the integrand of multi-loop amplitudes in planar $\mathcal{N}=4$
SYM \cite{ArkaniHamed:2010kv}, see also \cite{Boels:2010nw}.

Another interesting structure, that is going to be essential in this review, appeared in 2008, when
Bern, Carrasco and Johansson (BCJ) found a curious \textit{color-kinematic duality} for gauge theory amplitudes \cite{BCJ}. By means of this duality
they were able to write down new relations, reducing the number of independent color-ordered gauge theory amplitudes from $(n-2)!$, as
given by the Kleiss-Kuijf relations \cite{KK,DelDuca:1999rs}, down to $(n-3)!$. At this stage these \textit{BCJ relations}, and their
supersymmetric extension \cite{Sondergaard:2009za}, had not been proven for general $n$-point amplitudes. Interestingly, the first proof came from string theory
where Bjerrum-Bohr, Damgaard and Vanhove \cite{monodromy} and Stieberger \cite{Stieberger:2009hq} used monodromy to derive the $(n-3)!$ basis for color-ordered open string amplitudes.
In the field theory limit these \textit{monodromy relations} reduce exactly to Kleiss-Kuijf and BCJ
relations. Nearly one year passed before the BCJ relations were also proven from
pure field theory \cite{Feng:2010my}, see also \cite{Jia:2010nz}, and most recently investigations of similar structures appearing at loop level have been made \cite{BjerrumBohr:2011xe}.

In mid 2010 all the tools needed for the first purely field theoretical proof of the KLT relations were at hand
\cite{KLTproofShort,KLTproofLong}. In the process several new ways of writing the relations were found (and needed) along with
the introduction of a \textit{momentum kernel} which nicely captured all of these forms, including the one conjectured in \cite{Bern:1998sv,Abe:2005se}.
This momentum kernel turned out to have a lot of nice properties which was investigated in \cite{Skernel}
along with its extension to string theory, and made it much easier to handle and express KLT relations for general $n$ points. It was also realized that KLT relations could lead to pure gauge theory identities
when non-matching helicities were chosen \cite{NewId}, see \cite{Feng:2010br,Tye:2010kg,Elvang:2010kc} on the extension to
supersymmetric theories and its connection to $SU(8)$-violating gravity amplitudes. In addition, it became clear how closely related
the BCJ and KLT relations actually are, although they appear very different.

The review is structured as follows; in section \ref{KLTstring_sec} we show how to derive the $n$-point KLT relation from string theory.
This will be done in quite some detail since the original paper is a bit brief concerning some of the more technical issues. We will also arrive at a
more explicit $n$-point form that nicely incorporates the freedom one has in different ways of expressing the relation. In section \ref{ftlimit_sec} we look at its field theory limit and properties in this regime. We also review some field theoretical tools that will be needed later on,
and comment some more on the close connection between BCJ and KLT relations mentioned above. In section \ref{ft_proof_sec} we utilize all of the structures
presented in section \ref{ftlimit_sec} in order to give a field theoretical proof of the KLT relations. In section \ref{BCJ_sec} we review
the color-kinematic duality approach and its connection to gravity, both at tree- and loop-level.
Finally, in section \ref{conclusions} we present our conclusions.

\section{Factorization of Closed String Amplitudes\label{KLTstring_sec}}
We begin the review with a derivation of the KLT relations from string theory. We take the same path as in the original paper \cite{KLT}
and in \cite{Skernel},
factorizing the closed string into the product of open strings -- glued together by appropriate phase factors.

The $n$-point tree-level closed string amplitude is given by
\begin{equation}
  \label{Mclosed}
  \mathcal{M}_n=\left(i\over 2\pi\alpha'\right)^{n-3}\,
\!\!\!\!\int \prod_{i=2}^{n-2} d^2z_i |z_i|^{2\alpha' k_1\cdot k_i} |z_i-1|^{2\alpha'\,k_{n-1}\cdot k_i}\!\!\!\!
\prod_{i<j\leq
    n-2}\!\! |z_j-z_i|^{2\alpha'\,k_i\cdot k_j}\, f(z_i)\,g(\bar z_i)\, ,
\end{equation}
where we have fixed the three points $z_1=0$, $z_{n-1}=1$ and  $z_n=\infty$. The $f(z_i)$ and $g(\bar{z}_i)$ functions
come from the operator product expansion of vertex operators and
depend on the type of external states we are considering, but since they are without any branch cuts
they will not be important for the following argument. We will denote $z_i = v_i^1 + i v_i^2$, such that
\begin{align}
& |z_i|^{2\alpha' k_1\cdot k_i} = \left[ (v_i^1)^2 + (v_i^2)^2\right]^{\alpha' k_1\cdot k_i}\,, \\
& |z_i-1|^{2\alpha'\,k_{n-1}\cdot k_i} = \left[ (v_i^1-1)^2 + (v_i^2)^2 \right]^{\alpha'\,k_{n-1}\cdot k_i}\,, \\
& |z_j-z_i|^{2\alpha'\,k_i\cdot k_j} = \left[ (v_j^1-v_i^1)^2 + (v_j^2-v_i^2)^2 \right]^{\alpha'\,k_i\cdot k_j}.
\end{align}
By analytically continuing the $v_i^2$ variables to the complex plane,
we can rotate the integration contour for these variables from the real axis to (almost) the imaginary axis
\begin{align}
v_i^2 \quad \longrightarrow \quad ie^{-2i\epsilon}v_i^2 \simeq  i(1-2i\epsilon)v_i^2\,,
\end{align}
without changing the value of the amplitude. Here $\epsilon >0$ is some small number making sure we avoid the branch points.
This changes the expressions in the integrand (to linear order in $\epsilon$)
\begin{align}
 \left[ (v_i^1)^2 + (v_i^2)^2\right]^{\alpha' k_1\cdot k_i} & \longrightarrow \left[ (v_i^1)^2 - (v_i^2)^2
+ 4i\epsilon (v_i^2)^2 \right]^{\alpha' k_1\cdot k_i} \,, \label{tranOne}  \\
\left[ (v_i^1-1)^2 + (v_i^2)^2 \right]^{\alpha'\,k_{n-1}\cdot k_i} & \longrightarrow
\left[ (v_i^1)^2 - (v_i^2)^2 -2v_i^1 + 1 + 4i\epsilon (v_i^2)^2 \right]^{\alpha'\,k_{n-1}\cdot k_i} \label{tranTwo} \, , \\
\left[ (v_j^1-v_i^1)^2 + (v_j^2-v_i^2)^2 \right]^{\alpha'\,k_i\cdot k_j} & \longrightarrow
\left[ (v_j^1-v_i^1)^2 - (v_j^2 - v_i^2)^2(1 - 4i\epsilon ) \right]^{\alpha'\,k_i\cdot k_j} \label{tranThree} \, .
\end{align}
If we now make the transformation of variables
\begin{align}
v_i^\pm \equiv  v_i^1\pm v_i^2 \,,
\end{align}
and define $\delta_i \equiv v_i^+-v_i^-$, it is easy to verify that the expression on the right-hand-side of line \eqref{tranOne},
\eqref{tranTwo} and \eqref{tranThree} is given by
\begin{align}
& (v_i^+ -i\epsilon\delta_i)^{\alpha'k_1\cdot k_i} (v_i^- +i\epsilon\delta_i)^{\alpha'k_1\cdot k_i} \,, \quad
(v_i^+ -1 -i\epsilon\delta_i)^{\alpha'k_{n-1}\cdot k_i} (v_i^- -1 +i\epsilon\delta_i)^{\alpha'k_{n-1}\cdot k_i} \,,
\label{split_one}
\end{align}
and
\begin{align}
\big(v_i^+-v_j^+
-i\epsilon (\delta_i-\delta_j)\big)^{\alpha'\,k_i\cdot k_j}\big(v_i^--v_j^-+i\epsilon (\delta_i-\delta_j)\big)^{\alpha'\,k_i\cdot k_j} \,,
\label{split_two}
\end{align}
respectively. 

In total, this brings eq.~\eqref{Mclosed} into the form
\begin{align}
\label{almostFac}
&\mathcal{M}_n= \left(\frac{i}{2}\right)^{n-3}\left(i\over 2\pi\alpha'\right)^{n-3}\!\!
\int_{-\infty}^{+\infty} \prod_{i=2}^{n-2} dv_i^+dv_i^-
f(v^-_i)\,g(v_i^+)\nonumber \\
&\hspace{1cm}\times
(v_i^+-i\epsilon\delta_i)^{\alpha' k_1\cdot k_i}
(v_i^-+i\epsilon\delta_i)^{\alpha' k_1\cdot k_i}
(v_i^+-1-i\epsilon\delta_i)^{\alpha'\,k_{n-1}\cdot k_i}
(v_i^--1+i\epsilon\delta_i)^{\alpha'\,k_{n-1}\cdot k_i}\nonumber \\
&\hspace{1cm}\times \prod_{i<j\leq n-2} \big(v_i^+-v_j^+
-i\epsilon (\delta_i-\delta_j)\big)^{\alpha'\,k_i\cdot k_j}
\big(v_i^--v_j^-+i\epsilon (\delta_i-\delta_j)\big)^{\alpha'\,k_i\cdot k_j}\,,
\end{align}
where the additional factor of $(i/2)^{n-3}$ is due to the Jacobian when changing variables in the integral and from the rotation of
the $v_i^2$ contours.

First we note the following; assume that at least one $v_i^+ \in\, ]-\infty,0[\,$ and look at the contribution from $v_i^-$, \textit{i.e.}
\begin{align}
& \int_{-\infty}^{+\infty} dv_i^-
f(v^-_i)(v_i^-+i\epsilon\delta_i)^{\alpha' k_1\cdot k_i}
(v_i^--1+i\epsilon\delta_i)^{\alpha'\,k_{n-1}\cdot k_i} \!\!\! \prod_{i<j\leq n-2} \!\!\!
\big(v_i^--v_j^-+i\epsilon (\delta_i-\delta_j)\big)^{\alpha'\,k_i\cdot k_j}\,.
\end{align}
The behaviour of the imaginary $\epsilon$ terms near the branch points is
\begin{align}
v_i^- \sim  0 \quad & \Longrightarrow \quad \delta_i \sim v_i^+ < 0 \,, \nonumber \\
v_i^- \sim  1 \quad & \Longrightarrow \quad \delta_i \sim v_i^+ - 1 < 0 \,, \nonumber \\
v_i^- \sim  v_j^- \quad & \Longrightarrow \quad \delta_i-\delta_j \sim v_i^+ - v_j^+ < 0 \quad \mathrm{when} \quad v_i^+ < v_j^+.
\end{align}
The requirement $v_i^+ < v_j^+$ in the last line is not a problem, since we can just choose to look at the $v_i^-$ integral corresponding to the
``smallest'' $v_i^+$ variable, which has to lie in the range $]-\infty,0[\,$ due to our first assumption.
This means that we can close the integral of $v_i^-$ in the lower half of the complex $v_i^-$-plane (again by analytical continuation),
and since the closed contour does not contain any poles the integral vanishes. In general, when $v_i^+ < v_j^+$ we avoid the branch point
$v_i^- = v_j^-$ below the real axis, and when $v_i^+ > v_j^+$ we avoid it above the real axis. From this kind of
argument we see that whenever one of the $v_i^+$-variables is in the range of $]-\infty, 0[$ or $]1,\infty[\,$, at least one of the $v_i^-$ contours
can be completely closed either below or above the real axis.
Hence, only when all $v_i^+$ lie between 0 and 1 will there be a contribution to eq.~\eqref{almostFac}.

By splitting up the $v_i^+$-integration region we can write the $n$-point closed string amplitude as
\begin{equation}\label{Msum}
\mathcal{M}_n=\sum_{\sigma}          \,
M_n^{\sigma}\big(\sigma(2),\ldots,\sigma(n-2)\big)\,,
\end{equation}
where $M_n^{\sigma}(\sigma(2),\cdots,\sigma(n-2))$ is the ``ordered amplitude'' defined such that $v^+_{\sigma(2)}    <
v^+_{\sigma(3)}<   \cdots    < v^+_{\sigma(n-2)}$. For instance, at five points this corresponds to splitting the integration over
the $(v_2^+,v_3^+)$-plane into an integral over the region ``above'' the $v_3^+ = v_2^+$ line (\textit{i.e.} $v_2^+ < v_3^+$)
and an integral ``below'' this line (\textit{i.e.} $v_3^+ < v_2^+$). Together with the above restriction on the $v_i^+$-integration range,
the $v_i^+$ part of $M_n^{\sigma}$ in eq.~\eqref{Msum} becomes
\begin{align}
   \int_{ 0<v^+_{\sigma(2)} <  \cdots <
v^+_{\sigma(n-2)}<1}       \prod_{i=2}^{n-2}        dv_{i}^+\,
  g\big(v^+_{i}\big)
  \big(v_{i}^+\big)^{\alpha' k_1\cdot k_{i}}
\big(1-v_{i}^+\big)^{\alpha'\,k_{n-1}\cdot k_{i}} \nonumber \\
& \hspace{-5cm} \times
\prod_{{i}<{j}\leq
    n-2}\big(v_{\sigma(j)}^+-v_{\sigma(i)}^+\big)^{\alpha'\,k_{\sigma(i)}\cdot k_{\sigma(j)}}\,,
\label{e:ALfinal}\end{align}
where we have omitted the infinitesimal $\epsilon$ terms. 
We recognize \eqref{e:ALfinal} as the expression exactly corresponding to the color-ordered open string
amplitude $\mathcal{A}_n(1,\sigma(2,\dots,n-2),n-1,n)$. Note that, compared to eq.~\eqref{almostFac}, we have
written $(1-v_i^+)^{\alpha'\,k_{n-1}\cdot k_{i}}$ instead of $(v_i^+-1)^{\alpha'\,k_{n-1}\cdot k_{i}}$ and
$(v_{j}^+-v_{i}^+)^{\alpha'\,k_{i}\cdot k_{j}}$ such that $v_j^+-v_i^+>0$ always. This is needed in order to
make the identification with a color-ordered open string amplitude, however, we are of course only allowed to do this if
we make a similar change in the $v_i^-$ part, otherwise we would get wrong phase factors.

For simplicity we will from now on fix the ordering to $\{ 2,3,\ldots,n-2\}$, \textit{i.e.} we are considering the $M_n^{\sigma}(2,3,\ldots,n-2)$
contribution in eq.~\eqref{Msum}. The remaining terms can simply be obtained through permutation of labels.

\begin{figure}[t]
\centering
\includegraphics[width=12cm]{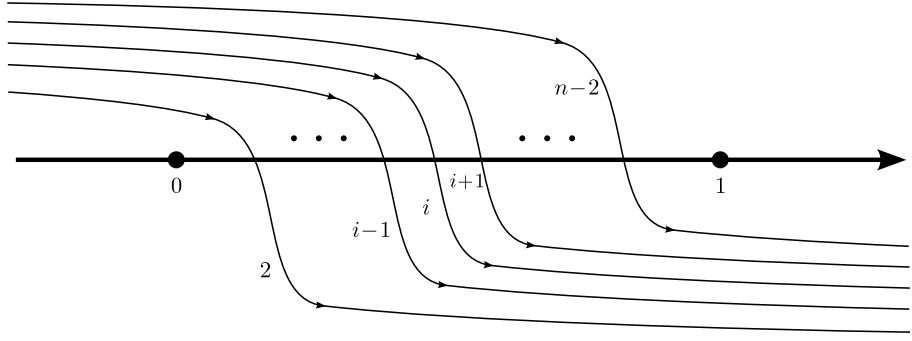}
\caption{\sl The nested structure of the contours of integration
for the $v^-_i$ variables   corresponding       to        the       ordering
$0<v^+_2<v^+_3<\cdots<v^+_{n-2}<1$ of the $v_i^+$ variables.}
\label{fig:contourNested}
\end{figure}

We have just seen that the $v_i^+$ part is nothing but the color-ordered $\mathcal{A}_n(1,2,\dots,n)$ amplitude. We therefore turn our attention to the $v_i^-$ part, investigating which contours the imaginary $\epsilon$ terms
dictate for the integrals. Near $v_i^-\sim 0$ the quantity $i\epsilon \delta_i\sim
i\epsilon v_i^+$ is a positive imaginary number (remember that $v_i^+\in ]0,1[\,$), so
the contour is above the real axis here. For $v_i^-\sim 1$ we have $i\epsilon \delta_i \sim i\epsilon
(v_i^+-1)$ which is a negative imaginary number, hence the contour
lies below the real axis.  Finally, for $v^-_i\sim v^-_j$ we see that $i\epsilon(\delta_i-\delta_j) \sim i\epsilon(v_i^+ - v_j^+)$,
meaning that the contour for $v^-_i+i\epsilon \delta_i$ lies below the contour of $v^-_j+i\epsilon \delta_j$ for $i<j$. See figure~\ref{fig:contourNested} for an illustration of this
nested structure.

%===========================================================================
%===========================================================================

The next step is to deform the contours for $v^-_i+i\epsilon \delta_i$ into expressions corresponding to color-ordered
amplitudes. That is, we are going to close the contours either to the left, turning the contour below the real axis,
or to the right, turning the contour above the real axis. Besides having the correct integration \textit{region}, in order to identify the
integrals with amplitudes, we also need to make sure that the \textit{integrand} is correct. This implies that we sometime
need to pull out phase factors. However, in order to not cross a branch cut
we do this in the following way; for $z^c$ with $\mathrm{Re}(z)<0$
\begin{equation}
z^c= \begin{cases}
e^{i\pi c} (-z)^c& {\rm Im} (z)\geq 0\,,\cr
e^{-i\pi c}\, (-z)^c& {\rm Im} (z)<0\,.
\end{cases}
\label{phase_factors}
\end{equation}
We have added some additional details concerning this subtle, but important, point in appendix~\ref{branchcuts}.

Furthermore, there is a freedom in the number of contours
one can close to the left or the right. For a given
$2\leq j\leq n-1$, we can pull the contours from
2 up to $j-1$ to the left, and the set from $j$ to $n-2$ to
the right ($j=2$ or $j=n-1$ means \textit{all} to the right or \textit{all} to the left, respectively).
Let us illustrate this in details for the five point case (the four point case is a bit to simple in order to
capture all the features of the general argument.)

\subsection{Five Point KLT Relations}
\begin{figure}[t]
\centering
\includegraphics[width=12cm]{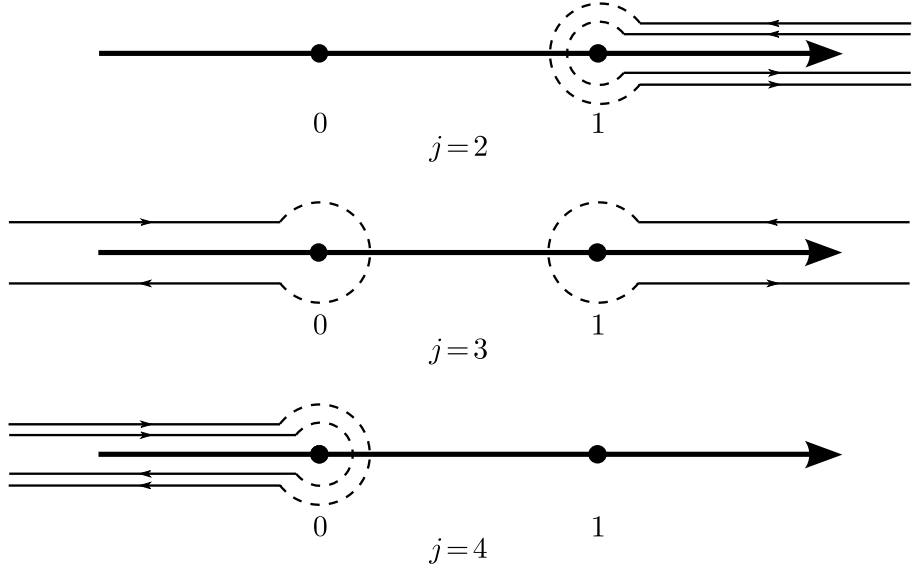}
\caption{\sl The three different ways of flipping contours in the five-point case.}
\label{fig:5ptleft}
\end{figure}

Starting with $j=4$, pulling the contour for $v_2^-$ to the left,
only showing the piece involving $v_2^-$, we get
\begin{align}
  \label{e:C2}
  &\int_{C_2}          dv_2^-  \,       (v_2^-)^{\alpha'k_1\cdot         k_2}
  (1-v_2^-)^{\alpha'k_{4}\cdot                  k_2}\,
  (v_3^--v_2^-)^{\alpha'k_3\cdot k_2}\,f(v_2^-) \nonumber \\
&=(e^{i\pi \alpha' k_1\cdot k_2}-e^{-i\pi \alpha' k_1\cdot k_2})
  \int_{-\infty}^0 dv_2^-\,        (-v_2^-)^{\alpha'k_1\cdot         k_2}
  (1-v_2^-)^{\alpha'k_{4}\cdot                  k_2}\,
  (v_3^--v_2^-)^{\alpha'k_3\cdot k_2}\,f(v_2^-) \nonumber \\
&=2i\sin(\pi \alpha' k_1\cdot k_2)
  \int_{-\infty}^0 dv_2^-\,        (-v_2^-)^{\alpha'k_1\cdot         k_2}
  (1-v_2^-)^{\alpha'k_{4}\cdot                  k_2}\,
  (v_3^--v_2^-)^{\alpha'k_3\cdot k_2}\,f(v_2^-)\,.
\end{align}
Note that in the first line above we write $(1-v_2^-)^{\alpha'k_{4}\cdot k_2}$ instead of $(v_2^--1)^{\alpha'k_{4}\cdot k_2}$
and $(v_3^--v_2^-)^{\alpha'k_3\cdot k_2}$ instead of $(v_2^--v_3^-)^{\alpha'k_3\cdot k_2}$
in order to compensate for the same swapping of order in the $v_i^+$ integral of eq.~\eqref{e:ALfinal}.
Now, as illustrated in the bottom of figure~\ref{fig:5ptleft}, we also close the contour for $v_3^-$ to the left (only showing the part involving
the $v_3^-$ variable)
\begin{align}
  \label{e:C3}
&\int_{C_3}          dv_3^-   \,      (v_3^-)^{\alpha'k_1\cdot k_3}
  (1-v_3^-)^{\alpha'k_{4}\cdot k_3}(v_3^--v_2^-)^{\alpha'k_3\cdot k_2}\,f(v_3^-) \nonumber \\
&= (e^{i\pi \alpha' (k_1+k_2)\cdot k_3}-e^{-i\pi \alpha' (k_1+k_2)\cdot k_3}) \,  \int_{-\infty}^{v_2^-} dv_3^-\,
(-v_3^-)^{\alpha'k_1\cdot         k_3}
  (1-v_3^-)^{\alpha'k_{4	}\cdot k_3}(v_2^--v_3^-)^{\alpha'k_2\cdot k_3}\,f(v_3^-) \nonumber \\
&\phantom{aa} +(e^{i\pi \alpha' k_1\cdot k_3}-e^{-i\pi \alpha' k_1\cdot k_3})
  \int_{v_2^-}^{0} dv_3^-\,     (-v_3^-)^{\alpha'k_1\cdot         k_3}
  (1-v_3^-)^{\alpha'k_{4}\cdot k_3}(v_3^--v_2^-)^{\alpha'k_2\cdot k_3}\,f(v_3^-) \nonumber \\
&= 2i\sin\big(\pi \alpha' (k_1+k_2)\cdot k_3\big)  \,  \int_{-\infty}^{v_2^-} dv_3^-\,
(-v_3^-)^{\alpha'k_1\cdot         k_3}
  (1-v_3^-)^{\alpha'k_{4	}\cdot k_3}(v_2^--v_3^-)^{\alpha'k_2\cdot k_3}\,f(v_3^-) \nonumber \\
&\phantom{aa} +2i\sin(\pi \alpha' k_1\cdot k_3)
  \int_{v_2^-}^{0} dv_3^-\,     (-v_3^-)^{\alpha'k_1\cdot         k_3}
  (1-v_3^-)^{\alpha'k_{4}\cdot k_3}(v_3^--v_2^-)^{\alpha'k_2\cdot k_3}\,f(v_3^-) \,.
\end{align}
We see that the total integration over $v_2^-$ and $v_3^-$ exactly correspond to color-ordered open string amplitudes,
which we will denote $\widetilde{\mathcal{A}}_5$ to distinguish them from the ones following from the $v_i^+$ part,
such that the whole $v_i^-$ contribution in $M_5^{\sigma}(2,3)$ can be written as
\begin{align}
\propto\,\, \sin(\pi \alpha' k_1\cdot k_2)\sin\big(\pi \alpha' (k_1+k_2)\cdot k_3\big) \mathcal{\widetilde{A}}_5(3,2,1,4,5) \nonumber \\
 +\sin(\pi \alpha' k_1\cdot k_2)\sin(\pi \alpha' k_1\cdot k_3) \mathcal{\widetilde{A}}_5(2,3,1,4,5)\,.
\end{align}
Together with eq.~\eqref{e:ALfinal} for $n=5$, and eq.~\eqref{Msum}, we have obtained the following relation between
the five-point closed string amplitude $\mathcal{M}_5$ and the color-ordered open string amplitudes $\mathcal{A}_5$, $\mathcal{\widetilde{A}}_5$
\begin{align}
\mathcal{M}_5 ={}& \frac{-1}{4\pi^2\alpha'^2} \left[ \sin(\pi \alpha' k_1\cdot k_2)\sin\big(\pi \alpha' (k_1+k_2)\cdot k_3\big) \mathcal{A}_5(1,2,3,4,5) \mathcal{\widetilde{A}}_5(3,2,1,4,5) \right. \nonumber \\
 & \hspace{2cm}\left.+\sin(\pi \alpha' k_1\cdot k_2)\sin(\pi \alpha' k_1\cdot k_3) \mathcal{A}_5(1,2,3,4,5)\mathcal{\widetilde{A}}_5(2,3,1,4,5) \right] \nonumber \\
& + (2\leftrightarrow 3) \,.
\label{Mj=4}
\end{align}
%
%===========================================================================
%===========================================================================
If we take the other extreme, \textit{i.e.} closing both contours to the right ($j=2$), we get
\begin{align}
  \label{e:Cn-2}
2i\sin(\pi \alpha' k_{4}\cdot k_{3})
  \int_{1}^{+\infty} dv_{3}^-\,
(v^-_{3})^{\alpha'k_1\cdot         k_{3}}
  (v_{3}^--1)^{\alpha'k_{4}\cdot
k_{3}}  (v_{3}^--v_2^-)^{\alpha'k_2\cdot k_{3}}\,f(v_{3}^-)\,,
\end{align}
for the $v_3^-$ integration, and
\begin{align}
  \label{e:Cn-3}
& 2i\sin(\pi \alpha' k_{4}\cdot k_{2})
  \int_{1}^{v_{3}^-} dv_{2}^-\,(v_{2}^-)^{\alpha'k_1\cdot k_{2}}
  (v_{2}^--1)^{\alpha'k_{4}\cdot k_{2}} (v_{3}^--v_{2}^-)^{\alpha'k_{3}\cdot k_{2}}\,f(v_{2}^-)\\
\nn&+2i\sin\big(\pi \alpha' (k_{4}+k_{3})\cdot k_{2}\big) \,
\int_{v_{3}^-}^{+\infty} dv_{2}^-\,
(v_{2}^-)^{\alpha'k_{4}\cdot         k_{2}}
  (v_{2}^--1)^{\alpha'k_{4}\cdot k_{2}} (v_{2}^--v_{3}^-)^{\alpha'k_{3}\cdot k_{2}}\,f(v_{2}^-)\,,
\end{align}
for the $v_2^-$ integration, see the top case of figure~\ref{fig:5ptleft}, \textit{i.e.}
\begin{align}
\mathcal{M}_5 ={}& \frac{-1}{4\pi^2\alpha'^2} \left[ \sin(\pi \alpha' k_{4}\cdot k_{3})\sin(\pi \alpha' k_{4}\cdot k_{2}) \mathcal{A}_5(1,2,3,4,5) \mathcal{\widetilde{A}}_5(1,4,2,3,5) \right. \nonumber \\
 & \hspace{2cm}\left.+\sin(\pi \alpha' k_{4}\cdot k_{3})\sin\big(\pi \alpha' (k_{4}+k_{3})\cdot k_{2}\big)\mathcal{A}_5(1,2,3,4,5)\mathcal{\widetilde{A}}_5(1,4,3,2,5) \right] \nonumber \\
& + (2\leftrightarrow 3) \,.
\label{Mj=2}
\end{align}
Finally, we could have closed $v_2^-$ to the left and $v_3^-$ to the right, as also illustrated in
figure~\ref{fig:5ptleft}, resulting in ($j=3$)
\begin{align}
\mathcal{M}_5 ={}& \frac{-1}{4\pi^2\alpha'^2} \sin(\pi \alpha' k_{1}\cdot k_{2})\sin(\pi \alpha' k_{4}\cdot k_{3}) \mathcal{A}_5(1,2,3,4,5) \mathcal{\widetilde{A}}_5(2,1,4,3,5)  \nonumber \\
& + (2\leftrightarrow 3) \,.
\label{Mj=3}
\end{align}

All of these different forms can be nicely collected into one compact formula if we introduce the \textit{momentum kernel}
\begin{equation}
  \label{Sn}
  \mathcal{S}_{\alpha'}[i_1,\ldots,i_k|
j_1,\ldots,j_k]_{p} \equiv (\pi\alpha'/2)^{-k}\,
\prod_{t=1}^{k}\, \sin \big(\pi\alpha'\,(p\cdot
  k_{i_t}+ \sum_{q>t}^{k} \, \theta(i_t,i_q)\, k_{i_t}\cdot k_{i_q})  \big)\,,
\end{equation}
where $\theta(i_t,i_q)$ equals 1 if the ordering of
$i_t$ and $i_q$ is opposite in $\{i_1,\ldots,i_k\}$
and $\{j_1,\ldots,j_k\}$, and 0 if the ordering is the same.
Here we have defined $\mathcal{S}_{\alpha'}$ for a general number of legs, so, for instance
\begin{align}
\mathcal{S}_{\alpha'}[2|2]_{k_1} ={}& (\pi\alpha'/2)^{-1} \sin \big(\pi\alpha' k_1\cdot k_{2}  \big)\,,  \nonumber \\
\mathcal{S}_{\alpha'}[23|23]_{k_1} ={}& (\pi\alpha'/2)^{-2} \sin \big(\pi\alpha' k_1\cdot k_{2}  \big)
\sin \big(\pi\alpha' k_1\cdot k_{3}  \big)\,, \nonumber  \\
\mathcal{S}_{\alpha'}[23|32]_{k_1} ={}& (\pi\alpha'/2)^{-2} \sin \big(\pi\alpha' (k_1+k_3)\cdot k_{2}  \big)
\sin \big(\pi\alpha' k_1\cdot k_{3}  \big)\,,
\end{align}
and so on. We will also define $\mathcal{S}_{\alpha'}[\emptyset|\emptyset]_p = 1$ for empty sets.
With this $\mathcal{S}_{\alpha'}$ function we can collect eq.~\eqref{Mj=4}, \eqref{Mj=2} and \eqref{Mj=3} into one formula as
\begin{align}
 \mathcal{M}_5 ={}& \left(-i/4\right)^{2} \times \nonumber \\
&\hspace{-1.2cm}\sum_{\sigma}
\sum_{\gamma,\beta}
\mathcal{S}_{\alpha'}[\gamma(\sigma(2),\dots,\sigma(j\!-\!1))|\sigma(2),\dots,\sigma(j\!-\!1)]_{k_1}
\mathcal{S}_{\alpha'}[\sigma(j),\dots,\sigma(3)|
\beta(\sigma(j),\dots,\sigma(3))]_{k_{4}}\nonumber \\
& \times
   \cA_5(1,\sigma(2,3),4,5)\,
   \widetilde{\cA}_5(\gamma(\sigma(2),\dots,\sigma(j\!-\! 1)),1,4,\beta(\sigma(j),\dots,\sigma(3)),5)\,,
\label{M5final}
\end{align}
with $j=\{2,3,4\}$. Note that when $j=2$ one should read the set $\{\sigma(2),\dots,\sigma(j\!-\!1)\}$ as being empty,
and likewise for $j=4$ the set $\{ \sigma(j),\ldots,\sigma(3)\}$ is empty.

\subsection{General $n$-point KLT Relations}

Eq.~\eqref{M5final} not only collects all the different five point forms, due to different ways of closing the contours, in one nice expression,
but going through the same procedure for the $n$-point case, it directly generalizes to
\begin{align}
 \mathcal{M}_n ={}& \left(-i/4\right)^{n-3}\times \nonumber \\&\hspace{-1.2cm}\sum_{\sigma}
\sum_{\gamma,\beta}
\mathcal{S}_{\alpha'}[\gamma(\sigma(2),\dots,\sigma(j\!-\!1))|\sigma(2,\dots,j\!-\!1)]_{k_1}
\mathcal{S}_{\alpha'}[\sigma(j,\dots,n\!-\! 2)|
\beta(\sigma(j),\dots,\sigma(n\!-\! 2))]_{k_{n\!-\!1}}\nonumber \\
&\hspace{-0.9cm} \times
   \cA_n(1,\sigma(2,\dots,n\!-\! 2),n\!-\! 1,n)\,
   \widetilde{\cA}_n(\gamma(\sigma(2),\dots,\sigma(j\!-\! 1)),1,n\!-\!1,\beta(\sigma(j),\dots,\sigma(n\!-\!2)),n)\,,
  \label{Mnfinal}
\end{align}
with $2\leq j \leq n-1$.

Expression~\eqref{Mnfinal} shows how to factorize an $n$-point closed string amplitude $\mathcal{M}_n$ into the product of
$n$-point color-ordered open string amplitudes $\mathcal{A}_n$ and $\mathcal{\widetilde{A}}_n$, ``glued'' together by
kinematic factors contained in the $\mathcal{S}_{\alpha'}$ function. We note that
the expression is a sum over $(n-3)!\times
(j-2)!\times (n-1-j)!$ terms, taking its maximum value
$(n-3)!\times (n-3)!$ for  $j=2$  or  $j=n-1$, and its minimum $(n-3)!\times(\left\lceil {n\over2}\right\rceil -2)!\times (\left\lfloor {n\over2}\right\rfloor-1)!$ for $j=\lceil n/2\rceil$\footnote{The  floor  and  ceiling
  functions are defined on half-integers as follows:
  $\lfloor n/2\rfloor=(n-1)/2$ if $n$ is odd, or $n/2$ if
$n$ is even. $\lceil n/2\rceil=(n+1)/2$ if $n$ is odd, or $n/2$ if
$n$ is even.}.

Although, being one of the expressions having most terms, choosing $j=n-1$ the relation takes a particularly nice $n$-point form
\begin{align}
\label{stringPureKLTn}
\mathcal{M}_n &=\left(-i/4\right)^{n-3} \sum_{\sigma,\gamma}
\mathcal{S}_{\alpha'}[\gamma(2,\dots,n-2)|\sigma(2,\dots,n-2)]_{k_1}\cr
&\hspace{1cm}\times \cA_n(1,\sigma(2,\dots,n-2),n-1,n)
\widetilde{\cA}_n(n-1,n,\gamma(2,\dots,n-2),1)\,,
\end{align}
only involving one $\mathcal{S}_{\alpha'}$ function and having a greater symmetry between the sums over different ordered $\mathcal{A}_n$
and $\mathcal{\widetilde{A}}_n$ amplitudes. An equally nice form occurs with $j=2$.

\section{The Field Theory Limit\label{ftlimit_sec}}
Now that we have obtained the KLT relations in string theory, we will take a closer look at the field theory limit of eq.~\eqref{Mnfinal},
\textit{i.e.} letting $\alpha' \rightarrow 0$. The amplitudes just go to their corresponding field theory expressions
\begin{align}
\mathcal{M}_n \longrightarrow M_n\,, \qquad
\mathcal{A}_n \longrightarrow A_n\,, \qquad
\mathcal{\widetilde{A}}_n \longrightarrow \widetilde{A}_n\,,
\end{align}
and $\mathcal{S}_{\alpha'} \longrightarrow \mathcal{S}_{0} \equiv \mathcal{S}$, where it follows from eq.~\eqref{Sn} that
\begin{align}
  \mathcal{S}[i_1,\ldots,i_k|
j_1,\ldots,j_k]_{p} = 
\prod_{t=1}^{k} \big( s_{p i_t} + \sum_{q>t}^{k} \, \theta(i_t,i_q)\, s_{{i_t}{i_q}} \big) \,,
  \label{Sn_ft}
\end{align}
with $s_{ij} \equiv (k_i + k_j)^2 = 2 k_i\cdot k_j$, or more generally $s_{ij\ldots k} = (p_i + p_j + \cdots + p_k)^2$, which will
be used later on. Let us at this point also make a slight change in the overall constant such that it fits with a more commonly used
normalization of field theory amplitudes.
The $n$-point KLT relations in field theory then take the final form
\begin{align}
 M_n ={}& (-1)^{n+1}\times \nonumber \\&\hspace{-1.2cm}\sum_{\sigma}
\sum_{\gamma,\beta}
\mathcal{S}[\gamma(\sigma(2),\dots,\sigma(j\!-\!1))|\sigma(2,\dots,j\!-\!1)]_{k_1}
\mathcal{S}[\sigma(j),\dots,\sigma(n\!-\! 2)|\beta(\sigma(j,\dots,n\!-\! 2))]_{k_{n\!-\!1}}\nonumber \\
&\hspace{-0.9cm} \times
   A_n(1,\sigma(2,\dots,n\!-\! 2),n\!-\! 1,n)\,
   \widetilde{A}_n(\gamma(\sigma(2),\dots,\sigma(j\!-\! 1)),1,n\!-\!1,\beta(\sigma(j),\dots,\sigma(n\!-\!2)),n)\,,
  \label{Mnfinal_ft}
\end{align}
again with the freedom $2\leq j \leq n-1$. We have not been very explicit about which theories the amplitudes belong to, and as one might have
anticipated from the string theory derivation they are actually valid for very general classes of amplitudes. We could, for instance, have an $\mathcal{N}=8$ supergravity
amplitude on the l.h.s. and $\mathcal{N}=4$ SYM amplitudes on the r.h.s., or even express the full tree-level gauge theory amplitude as products between
color-ordered gauge theory and a color-scalar theory \cite{Bern:1999bx,Du:2011js}. However, for simplicity, we will for the rest of this review just consider it as a relation between pure graviton and pure gluon amplitudes.

\begin{figure}[t]
\centering
\includegraphics[width=12cm]{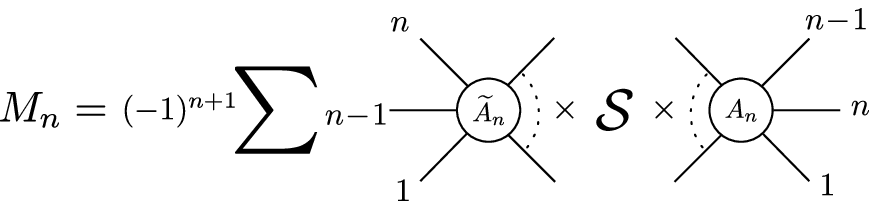}
\caption{\sl Diagrammatic representation of the KLT relation with $j=n-1$.}
\label{fig:KLT}
\end{figure}

Since it will be relevant for a later section, and because of its simple expression, let us explicitly write out the forms one obtain with $j=n-1$ and $j=2$, and at the same time
introduce a short-hand notation that will be used when expressions get large. With $j=n-1$ or $j=2$ eq.~\eqref{Mnfinal_ft} becomes
\begin{align}
M_n = (-1)^{n+1} \!\!\! \sum_{\sigma,\widetilde{\sigma} \in S_{n-3}} \!\!\!
\widetilde{A}_n(n-1,n,\widetilde{\sigma}_{2,n-2},1)
\mathcal{S}[\widetilde{\sigma}_{2,n-2}|\sigma_{2,n-2}]_{k_1}
A_n(1,\sigma_{2,n-2},n-1,n)\,,
\label{pureKLT}
\end{align}
or
\begin{align}
M_n = (-1)^{n+1} \!\!\! \sum_{\sigma,\widetilde{\sigma} \in S_{n-3}} \!\!\!
A_n(1,\sigma_{2,n-2},n-1,n)
\mathcal{S}[\sigma_{2,n-2}|\widetilde{\sigma}_{2,n-2}]_{k_{n-1}}
\widetilde{A}_n(1,n-1,\widetilde{\sigma}_{2,n-2},n)\,,
\label{dualKLT}
\end{align}
respectively, see also figure~\ref{fig:KLT}. Here we have introduced the short-hand notation $\sigma_{2,n-2} \equiv \sigma(2,\ldots,n-2)$, for a $\sigma$ permutation
over leg $\{2,\ldots,n-2\}$, and likewise for $\widetilde{\sigma}$. Note that $\sigma$ and $\widetilde{\sigma}$ are
two unrelated permutations, the tilde is just to remind us which one is going with $\widetilde{A}_n$,
and we are trying to be economic with our use of greek letters.

These relations are totally crossing symmetric, which they of course have to be since they equal a gravity amplitude. However, the r.h.s. is only
\textit{manifest} crossing symmetric in legs $ 2, 3,\ldots,n-2$. The crossing symmetry between, for instance legs $n$ and $n-1$ in eq.~\eqref{pureKLT} can also easily be seen by use of the reflection symmetry of color-ordered amplitudes $A_n(1,2,\ldots,n) = (-1)^nA_n(n,n-1,\ldots,1)$, and the following
identity
\begin{align}
  \mathcal{S}[i_1,\ldots,i_k|j_1,\ldots,j_k]_{p} =   \mathcal{S}[j_k,\ldots,j_1| i_k,\ldots,i_1]_{p}\,,
\label{STS}
\end{align}
but in general the crossing symmetry is not obvious at all. These comments also apply in the string theory case.

%===========================================================================
%===========================================================================

Although the calculations in section~\ref{KLTstring_sec} were a bit involved, at least one had 
an intuitive picture of breaking up a closed string into two open strings glued together by phase factors.
It was also clear how one could get different
expressions for this factorization by choosing different closures for the contours. The field theory limit then
follows naturally when taking $\alpha' \rightarrow 0$. However, it would be quite unsatisfactory if we could not
understand this field theory expression without going through string theory first. This will be our main focus for the remaining of this review;
how to see the KLT relations from a purely field theoretical point of view, including the possibility of going between
different expressions without having contours to deform. In order to get a better feel for the task that lies ahead let us
start by looking at some explicit lower-point examples and make some comments.

\subsection{Lower-point Examples}
The connection between gravity and gauge theory already starts at three points. It is well known that for real momenta
an on-shell three-point amplitude must vanish. However, if we go to complex momenta this is no longer the case. Interestingly, it
turns out that Lorentz invariance, and the spin of the external particles, uniquely fix the three-point amplitude, see \textit{e.g.} \cite{Paolo:2007}.
In detail, the spin 2 three-point amplitude (gravitons) is the product of two spin 1 three-point amplitudes (gluons)
\begin{align}
M_3(1,2,3) = A_3(1,2,3)\widetilde{A}_3(1,2,3)\,.
\label{3ptKLT}
\end{align}
This can also be directly seen from Feynman rules when legs are put on-shell, something that in particular simplifies the expression
for the gravity three-point vertex.
Although we did not have the complex three-point amplitude in mind when we wrote down eq.~\eqref{Mnfinal_ft}, it does give the
correct result even in this case. At first it might seem a bit strange to consider amplitudes with complex momenta, but recent
progress in amplitude calculations has largely been inspired by such an extension. Indeed, the BCFW recursion relation,
that will be reviewed below, relies on the deformation to complex momenta, and so will our proof of eq.~\eqref{Mnfinal_ft} in section \ref{ft_proof_sec}.

The four point case also looks very simple
\begin{align}
M_4(1,2,3,4) = - s_{12} A_4(1,2,3,4) \widetilde{A}_4(1,2,4,3)\,,
\label{4ptKLT}
\end{align}
however, its field theoretical origin already becomes a bit unclear here.
Compared to eq.~\eqref{3ptKLT} we see the appearance of a kinematic factor that makes sure to cancel one of the $s_{12}$ poles
present in both gauge theory amplitudes. We also see that the total crossing symmetry of the r.h.s. has already been well hidden, although
we know it has to be there.

When we go to five points, even expressed in the form with fewest terms ($j=3$),
\begin{align}
M_5 = s_{12}s_{34}A_5(1,2,3,4,5) \widetilde{A}_5(2,1,4,3,5) + s_{13}s_{24}A_5(1,3,2,4,5) \widetilde{A}_5(3,1,4,2,5)\,,
\end{align}
not only is the total crossing symmetry of the r.h.s. by no means obvious, but also the correct cancellation of poles begins to get more complicated
to see. These properties only get more and more disguised as we increase the number of external particles.

From the above examples it might seem almost impossible to identify, in all generality, the r.h.s. of eq.~\eqref{Mnfinal_ft}
with a gravity amplitude from a purely field theoretical/analytical point of view. Even the simplest features of gravity amplitudes
have become very non-trivial statements about the gauge theory side.
Before we can attack this problem we therefore need to review some important concepts and properties.
These will not only lead to a better understanding of KLT relations in field theory, but are by themselves amazing structures
of scattering amplitudes.
The first thing we will introduce is a rather unusual way of expressing the KLT relations. This form will have a higher degree of
manifest crossing symmetry, compared to eq.~\eqref{Mnfinal_ft}, but it requires a regularization. It has been proven separately by pure field
theory, and will turn out to be important even in the
proof of eq.~\eqref{Mnfinal_ft}.
Secondly, we introduce the BCJ relations, providing identities between color-ordered tree-level amplitudes,
and explain its connection to the $j$-independence of eq.~\eqref{Mnfinal_ft}.
Third, we look at what happens with the r.h.s. of eq.~\eqref{Mnfinal_ft} when $A_n$ and $\widetilde{A}_n$ belong
to different helicity sectors.
Fourth, we review the BCFW recursion relation, exploiting very general analytic properties of tree-level amplitudes to recursively
construct higher-point amplitudes from lower-point ones. All of these structures will be important for section \ref{ft_proof_sec},
where we will utilize them in order to give a purely field theoretical proof of eq.~\eqref{Mnfinal_ft}.

\subsection{Regularized KLT Relations}
Considering $n$-point amplitudes we start by making the following deformation of momenta $p_1$ and $p_n$
\begin{align}
p_1 \quad &\longrightarrow \quad p_1' = p_1 - xq \,, \nonumber \\
p_n \quad &\longrightarrow \quad p_n' = p_n + xq\,,
\end{align}
where $x$ is just some arbitrary parameter, and $q$ is a four-vector satisfying $q\cdot p_1 = q^2 =0$ and $q\cdot p_n \neq 0$.
This preserves conservation of momentum and keeps $p_1'^2 = 0$, but makes $p_n'^2 = s_{1'2\ldots (n-1)} \neq 0$.

The gravity amplitude $M_n$ can then be written as \cite{KLTproofShort}
\begin{align}
M_n=  (-1)^n \lim_{x\rightarrow 0} \sum_{\sigma,\widetilde{\sigma}\in S_{n-2}} \!\!\! \frac{\widetilde{A}_n(n',\widetilde{\sigma}_{2,n-1},1')
\mathcal{S}[ \widetilde{\sigma}_{2,n-1}|\sigma_{2,n-1}]_{p_1'} A_n(1',\sigma_{2,n-1},n')}{ s_{1'2\ldots(n-1)}}\,.
\label{newKLT}
\end{align}
We note that as $x\rightarrow 0$ the denominator goes to zero, but as we will see from eq.~\eqref{SA} below, so does the numerator. However, the whole
expression has a limit which is exactly equal to a gravity amplitude. This can also be seen by taking the soft limit of leg $n$
in eq.~\eqref{pureKLT} \cite{Skernel}, and comparing against the well known soft-limit behaviour of gravity amplitudes \cite{Bern:1998sv,Weinberg:1965nx}
(see \cite{Feng:2010hd} for an alternative approach.)
This might make eq.~\eqref{newKLT}
seem less strange, but we do stress that this expression for a gravity amplitude can be proven without knowledge of eq.~\eqref{pureKLT}
(or in general eq.~\eqref{Mnfinal_ft}).

If we make a deformation with $p_1'$ being off-shell instead of $p_n'$, we can write down the ``dual'' expression to eq.~\eqref{newKLT}
\begin{align}
M_n=(-1)^n \lim_{y\rightarrow 0} \sum_{\sigma,\widetilde{\sigma}\in S_{n-2}} \!\!\!
\frac{A_n(1',\sigma_{2,n-1},n')
\mathcal{S}[ \sigma_{2,n-1}| \widetilde{\sigma}_{2,n-1}]_{p_n'}
\widetilde{A}_n(n',\widetilde{\sigma}_{2,n-1},1')}{ s_{23\ldots n'}}\,,
\label{newKLTdual}
\end{align}
where we have called the deformation parameter $y$.

\subsection{BCJ Relations\label{BCJ_relations}}
The story of relations among color-ordered tree-level amplitudes is in itself very interesting. However, to not
lose focus of our current goal we will here only remind the reader about the
Bern-Carrasco-Johansson (BCJ) relations,
which will play an important role for us.

There are several different ways of presenting the BCJ relations, which all have their own advantages and disadvantages.
Here we choose one of the simplest general representations. For $n$-point amplitudes
these read
\begin{align}
0 ={}& s_{12} A_n(1,2,3,\ldots,n) + (s_{12} + s_{23}) A_n(1,3,2,4,\ldots,n) \nonumber \\
&+ (s_{12} + s_{23} + s_{24}) A_n(1,3,4,2,5,\ldots,n) + \cdots \nonumber \\
&+(s_{12} + s_{23} + s_{24} + \cdots + s_{2(n-1)}) A_n(1,3,4,\ldots,n-1,2,n)\,,
\label{BCJ}
\end{align}
along with all relations obtained by permutation of labels in the above expression. Note how leg 2 moves one position to the right
in each term and picks up an additional factor of $s_{2i}$, where $i$ is the leg just passed through. This particular form of the BCJ relations
actually contains all the information, in the sense that once you have these you can go from the $(n-2)!$ color-ordered amplitudes,
provided by the Kleiss-Kuijf relations, down to $(n-3)!$ independent amplitudes \cite{Feng:2010my}.

Some simple explicit examples of eq.~\eqref{BCJ} are
\begin{align}
0 = s_{12} A_4(1,2,3,4) + (s_{12} + s_{23}) A_4(1,3,2,4) \,,
\end{align}
for four points, and
\begin{align}
0 = s_{12} A_5(1,2,3,4,5) + (s_{12} + s_{23}) A_5(1,3,2,4,5) + (s_{12} + s_{23} + s_{24}) A_5(1,3,4,2,5)\,,
\end{align}
at five points. We could of course use momentum conservation to write some of the kinematic invariant factors simpler,
but keeping them like this make the relations easy to remember.

It turns out, that there is a very nice way of rephrasing the above BCJ relations in terms of our $\mathcal{S}$ function.
This is achieved through 
\begin{align}
0 = \sum_{\sigma \in S_{n-2}} \mathcal{S}[\beta(2,\ldots,n-1)|\sigma(2,\ldots,n-1)]_{k_1} A_n(1,\sigma(2,\ldots,n-1),n)\,,
\label{SA}
\end{align}
where $\beta$ is just some arbitrary permutation of the legs $\{2,3,\ldots,n-1\}$. This is nothing but a
sum of relations in the form of eq.~\eqref{BCJ}. To see this, let us first write out the five-point case with $\beta(2,3,4) = (2,3,4)$
explicitly, \textit{i.e.}
\begin{align}
& \sum_{\sigma \in S_{3}} \mathcal{S}[2,3,4|\sigma(2,3,4)]_{k_1} A_5(1,\sigma(2,3,4),5) \nonumber \\
&=  s_{13}s_{14} \big[ s_{12} A_5(1,2,3,4,5) + (s_{12}+s_{23})A_5(1,3,2,4,5) + (s_{12} + s_{23} + s_{24}) A_5(1,3,4,2,5) \big] \nonumber \\
&\phantom{aa} + s_{14}(s_{13}+s_{34}) \big[ s_{12} A_5(1,2,4,3,5) + (s_{12} + s_{24}) A_5(1,4,2,3,5) \nonumber \\
&\hspace{7.5cm} + (s_{12} + s_{23} + s_{24}) A_5(1,4,3,2,5) \big]\,.
\end{align}
The two $[\cdots]$ is exactly zero due to eq.~\eqref{BCJ}.

The general argument is to divide the sum of $\sigma(2,\ldots,n-1)$ into a
sum of groups where all, except the first leg in the $\beta$ ordering, call it $\beta(2)$, have fixed
ordering, and then insert $\beta(2)$ at any place. For each
group all factors from $\mathcal{S}$ will be the same except for the
factor contributing from $\beta(2)$.
This will give a BCJ relation and thereby vanish. In the above five-point example we had $\beta(2)=2$
and the two groups we summed over was the one with ordering $\{3,4\}$ and $\{4,3\}$, respectively.
From this we see that eq.~\eqref{SA} is a consequence of eq.~\eqref{BCJ}.

\subsubsection{The $j$-independence}
With the BCJ relations we can now address the $j$-independence of eq.~\eqref{Mnfinal_ft} from a field theory point of view, which
in string theory followed naturally due to the freedom in how one closes the contours. Indeed, it is useful to first establish the
equivalence between all forms only
differing in the $j$-value chosen. In this way, if we can prove eq.~\eqref{Mnfinal_ft} for just one specific form,
we have proven them all.

The form of BCJ relation that implies the $j$-independence is given by
\begin{align}
&\sum_{\alpha,\beta} \mathcal{S}[\alpha_{i_2,i_j}|i_2,\ldots,i_j]_{p_1}
\mathcal{S}[i_{j+1},\ldots,i_{n-2}|\beta_{i_{j+1},i_{n-2}}]_{p_{n-1}}
\widetilde{A}_n(\alpha_{i_2,i_j},1,n-1,\beta_{i_{j+1},i_{n-2}},n) \nonumber \\
&=\sum_{\alpha',\beta'} \mathcal{S}[\alpha'_{i_2,i_{j-1}}|i_2,\ldots,i_{j-1}]_{p_1}
\mathcal{S}[i_j,i_{j+1},\ldots,i_{n-2}|\beta'_{i_j,i_{n-2}}]_{p_{n-1}}
\widetilde{A}_n(\alpha'_{i_2,i_{j-1}},1,n-1,\beta'_{i_j,i_{n-2}},n)\,.
\label{jrel}
\end{align}
Although, eq.~\eqref{jrel} looks rather complicated it is straightforward to prove using only BCJ relations in the form of eq.~\eqref{BCJ},
and momentum conservation. Since the proof is not very enlightening we will not repeat it here, but simply refer to the appendix of \cite{KLTproofLong}
for details. Instead we will illustrate how to use eq.~\eqref{jrel} in the simple case of five points.

For $n=5$ and $j=2$ eq.~\eqref{jrel} reads (where we have taken $i_k = k$)
\begin{align}
\mathcal{S}[2|2]_{p_1}
\mathcal{S}[3|3]_{p_{4}}
\widetilde{A}_5(2,1,4,3,5) = \sum_{\beta'} \mathcal{S}[2,3|\beta'(2,3)]_{p_4}
\widetilde{A}_5(1,4,\beta'(2,3),5)\,,
\label{2rel_5pt}
\end{align}
and since eq.~\eqref{Mnfinal_ft} for $n=5$ and $j=2$ is given by
\begin{align}
 M_5 ={}& \sum_{\sigma}
\sum_{\beta}
   A_5(1,\sigma(2,3),4,5)\mathcal{S}[\sigma(2,3)|
\beta(2,3)]_{k_{4}}
   \widetilde{A}_5(1,4,\beta(2,3),5) \nonumber \\
={}&\sum_{\sigma}
   A_5(1,\sigma(2,3),4,5) \sum_{\beta}
\mathcal{S}[\sigma(2,3)|\beta(2,3)]_{k_{4}} \widetilde{A}_5(1,4,\beta(2,3),5) \,,
\end{align}
using eq.~\eqref{2rel_5pt} we immediately get
\begin{align}
M_5 ={}& \sum_{\sigma}
   A_5(1,\sigma(2,3),4,5)\mathcal{S}[\sigma(2)|\sigma(2)]_{p_1}
\mathcal{S}[\sigma(3)|\sigma(3)]_{p_{4}}
\widetilde{A}_5(\sigma(2),1,4,\sigma(3),5) \,,
\label{M5_jeq3}
\end{align}
which is just eq.~\eqref{Mnfinal_ft} with $j=3$. Likewise we could now use eq.~\eqref{jrel} with $j=3$ to
rewrite eq.~\eqref{M5_jeq3} into eq.~\eqref{Mnfinal_ft} with $j=4$.

This procedure generalizes to $n$ points; by repeated use of eq.~\eqref{jrel} the $j$-independence of eq.~\eqref{Mnfinal_ft} directly follows.

It is interesting to note that the BCJ relations could have been discovered much earlier.
Indeed, equating different expressions for the KLT relations lead to pure gauge theory amplitude relations, which, as we have just seen,
are directly related to the BCJ relations.
This was already apparent in \cite{KLT} but at that time not really appreciated. We also see that writing both the KLT and BCJ relations in terms of the
$\mathcal{S}$ function, their close connection becomes much more manifest than one would have anticipated from the original papers.

\subsection{Vanishing Identities\label{vanish_rel}}
The next surprising property of eq.~\eqref{Mnfinal_ft} we want to introduce is that if we take $A_n$ and $\widetilde{A}_n$
to live in different helicity \textit{sectors} the r.h.s. of eq.~\eqref{Mnfinal_ft} is identically zero.

To be more specific let us denote an $n$-point color-ordered tree amplitude with helicity configuration belonging to $N^kMHV$ as $A_n^k$,
\textit{i.e.} $A_n^k$ has $2+k$ negative helicity gluons, with $k\in\{0,1,\ldots,n-4\}$ for non-vanishing amplitudes.
We are not interested in the exact helicity configuration, just which helicity sector it belongs to. We then have \cite{NewId}
\begin{align}
 0={}& \sum_{\sigma}
\sum_{\gamma,\beta}
\mathcal{S}[\gamma(\sigma_{2,j\!-\!1})|\sigma_{2,j\!-\!1}]_{k_1}
\mathcal{S}[\sigma_{j,n\!-\! 2}|\beta(\sigma_{j,n\!-\! 2})]_{k_{n\!-\!1}} A_n^h(1,\sigma_{2,j\!-\!1},\sigma_{j,n\!-\! 2},n\!-\! 1,n)\nonumber \\
&\hspace{1.5cm}\times
   \widetilde{A}_n^{k\neq h}(\gamma(\sigma_{2,j\!-\! 1}),1,n\!-\!1,\beta(\sigma_{j,n\!-\!2}),n)\,,
\label{vanish}
\end{align}
which is nothing but the r.h.s. of eq.~\eqref{Mnfinal_ft}, written in our short-hand notation, with the color-ordered amplitudes
living in different helicity sectors.

At four points these relations are trivial, in the sense that we always get at least one amplitude that vanishes all by itself, however,
already at five points non-trivial cancellations start to occur. For instance, in the form with $j=3$, we have
\begin{align}
0 ={}& s_{12}s_{34}A_5(1^-,2^-,3^+,4^+,5^+) \widetilde{A}_5(2^-,1^-,4^+,3^-,5^+) \nonumber \\
& + s_{13}s_{24}A_5(1^-,3^+,2^-,4^+,5^+) \widetilde{A}_5(3^-,1^-,4^+,2^-,5^+)\,.
\end{align}

We also note that just as the KLT relations could be written in the form of eq.~\eqref{newKLT} and \eqref{newKLTdual} with a regularization,
so can these vanishing relations.
Eq.~\eqref{vanish} can be directly proven from the analytic properties of tree-level scattering amplitudes, but there is also
a more physical understanding of these identities. Looking at the KLT relations from an $\mathcal{N}=8$ supergravity point of view,
the vanishing of the r.h.s. of eq.~\eqref{Mnfinal_ft}, when $A_n$ and $\widetilde{A}_n$ belong to different helicity sectors,
correspond to $SU(8)$-violating gravity amplitudes and must therefore vanish \cite{Feng:2010br,Tye:2010kg,Elvang:2010kc} (see also \cite{Bianchi:2008pu}.)

\subsection{BCFW Recursion Relation}
Contrary to the properties and relations reviewed above, this section is not directly related to the KLT relations, but
will be essential for section \ref{ft_proof_sec}. Not only will the BCFW expansion of amplitudes in itself be important, but
also the method from which it can be derived \cite{BCFW}.

Start by deforming two of the external momenta, say $p_j$ and $p_l$, as
\begin{align}
p_j \quad &\longrightarrow \quad \widehat{p}_j = p_j - zq \,, \nonumber \\
p_l \quad &\longrightarrow \quad \widehat{p}_l = p_l + zq\,,
\end{align}
where $z$ is a complex variable and $q$ is a four-vector satisfying $q^2=q\cdot p_j = q \cdot p_l = 0$. This preserves
conservation of momentum and on-shellness. At tree-level $A_n$ is a rational function of external momenta, implying that
the deformed amplitude $A_n(z)$ is a rational function of $z$. Assuming $A_n(z) \rightarrow 0$ when $z\rightarrow \infty$, Cauchy's Theorem
tells us that
\begin{align}
0 = \frac{1}{2\pi i} \oint \frac{dz}{z} A_n(z) = A_n(0) + \sum_{\mathrm{poles}\,\, z_p\neq 0} \frac{\mathrm{Res}_p(A_n(z),z_p)}{z_p}\,,
\label{cauchy}
\end{align}
where $A_n(0)$ is the undeformed amplitude, coming from the residue of the $z=0$ pole.
The remaining poles (those different from $z=0$) come from $A_n(z)$, which we know can only follow from Feynman propagators going on-shell,
\textit{i.e.} when $P_{k,m}(z)^2$ vanishes, where
\begin{eqnarray}
P_{k,m}(z) \equiv p_k + p_{k+1} +\dots+\widehat{p}_j+\dots + p_m,
\end{eqnarray}
with $j\in \{k,\ldots,m\}$ and $l\notin \{k,\ldots,m\}$ (or vice versa). We do not get poles in $z$ if $j,l\in \{k,\ldots,m\}$ since
such a $P_{k,m}$ is independent of $z$.

Let us denote the value of $z$ where $P_{k,m}(z)$ is going on-shell by $z_{k,m}$, which can be found by solving $P_{k,m}(z_{k,m})^2=0$.
Since the poles of $A_n(z)$ will be \textit{simple poles}, the residues are given by
\begin{align}
\frac{\mathrm{Res}_p(A_n(z),z_{k,m})}{z_{k,m}} = -  \frac{\lim_{z\rightarrow z_{k,m}}[P_{k,m}(z)^2A_n(z)]}{P_{k,m}^2}\,,
\label{rescal}
\end{align}
where $P_{k,m} = P_{k,m}(0)$.
Using the general factorization property of gluon amplitudes
\begin{eqnarray}
A_n \xrightarrow{P^2\rightarrow 0}
\sum_{h=\pm}A_{r+1}(k,\ldots,m,-P^{-h})\frac{1}{P^2}A_{n-r+1}(P^{h},m+1,\ldots,k-1)\,,
\label{facgluon}
\end{eqnarray}
we see that
\begin{align}
\frac{\mathrm{Res}_p(A_n(z),z_{k,m})}{z_{k,m}} =
-\sum_{h=\pm}A_{r+1}(k,\ldots,m,-\widehat{P}_{k,m}^{-h})\frac{1}{P_{k,m}^2}A_{n-r+1}(\widehat{P}_{k,m}^{h},m+1,\ldots,k-1)\,,
\end{align}
with $\widehat{P}_{k,m} \equiv P_{k,m}(z_{k,m})$. Hence, combined with eq.~\eqref{cauchy}, we get the BCFW recursion relation \cite{BCF,BCFW}
\begin{eqnarray}
A_n = \sum_{r}\sum_{h=\pm}A_{r+1}(k,\ldots,m,-\widehat{P}_{k,m}^{-h})\frac{1}{P_{k,m}^2}A_{n-r+1}(\widehat{P}_{k,m}^{h},m+1,\ldots,k-1)\,,
\label{BCFW}
\end{eqnarray}
where the sum, denoted with $r$, states that we must sum over all internal momenta affected by the deformation.

This derivation was performed for color-ordered tree amplitudes, however, the recursion relation is also valid for tree-level gravity
amplitudes, we just need to sum over \textit{all} different combinations of momenta where the pole includes one of the deformed legs \cite{Bedford:2005yy}. It has been shown that the fall-off at $z \to
\infty$ for the graviton amplitude is even stronger
than one could naively have guessed \cite{zscaling,zscaling2}.

\section{A Purely Field Theoretical View\label{ft_proof_sec}}
We are finally in a position to see why eq.~\eqref{Mnfinal_ft} must be true for all $n$ from a purely field theoretical point of view.
This will be obtained in terms of an induction proof, and since we have already shown that the r.h.s. of
eq.~\eqref{Mnfinal_ft} is equivalent for all $j$-values, we are free to choose any of the versions we like the most. For us this
will be the ones shown in eq.~\eqref{pureKLT} and \eqref{dualKLT}.

\subsection{An $n$-point Proof}
We assume that we have checked eq.~\eqref{Mnfinal_ft} up to $n-1$ points, \textit{i.e.} we have checked that the expression on the r.h.s.
is equal to gravity. Then we write down the $n$-point expression for the r.h.s., let us denote it $R_n$.
Our goal is to show, only based on our knowledge of lower point cases, that this is equal to the $n$-point gravity amplitude,
that is $R_n = M_n$.

In the same way as the BCFW recursion relation was derived, we start out by deforming two momenta in our expression for $R_n$ and consider the contour integral
\begin{align}
B = \frac{1}{2\pi i}\oint \frac{dz}{z}R_n(z) = R_n(0) + \sum_{\mathrm{poles}\,\, z_p\neq 0} \frac{\mathrm{Res}_p(R_n(z),z_p)}{z_p}\,,
\label{cauchyR}
\end{align}
where $R_n(0)$ is just the undeformed $n$-point expression and we have included a potential boundary term $B$ on the left-hand side. Let us first argue that $B=0$. If we make a deformation in $p_1$ and $p_n$ the $\mathcal{S}$ function in eq.~\eqref{dualKLT} will be independent of $z$. The vanishing of the
boundary term is then guaranteed by the large $z$ behaviour of the gauge theory amplitudes $A_n$ and $\widetilde{A}_n$, this was also used in \cite{Du:2011js}. In the proof
we will present below it is more convenient to make a deformation in $p_1$ and $p_{n-1}$. However, such a deformation can not have a boundary term either
since we in section~\ref{ftlimit_sec} already argued for the crossing symmetry between $p_n$ and $p_{n-1}$ in $R_n$. The large $z$ behaviour must therefore be equally good for this deformation.

Now that we have established $B=0$, the goal is to show that the sum of residues exactly make up the
BCFW-expansion of an $n$-point gravity amplitude, and hence
\begin{align}
M_n \sim \sum_{\mathrm{poles}\,\, z_p\neq 0} \frac{\mathrm{Res}_p(R_n(z),z_p)}{z_p}\,.
\end{align}
If we can show this, using only lower-point cases and the properties/relations
reviewed in the last section, we are done. 

We begin by considering the residues that follow from poles of
the form $s_{12\ldots k}$, and we will be using $R_n$ in the form of eq.~\eqref{pureKLT}. Like in eq.~\eqref{rescal} we can can compute these from $\lim_{z\rightarrow
z_{12\ldots k}} \big[ s_{\widehat{1}2\ldots k}(z) R_n(z)
\big]/s_{12\ldots k}$, where $z_{12\ldots k}$ is the $z$-value
that makes $s_{\widehat{1}2\ldots k}$ go on-shell. 

There are basically only two classes of terms which have the possibility of contributing to a residue
\begin{itemize}
 \item (A) The pole appears only in \textit{one} of the amplitudes $\widetilde{A}_n$ or $A_n$.
 \item (B) The pole appears in \textit{both} $\widetilde{A}_n$ and $A_n$.
\end{itemize}
First we investigate (A), and we will consider the case where the pole only appears in $\widetilde{A}_n$. The case with
the pole appearing in $A_n$ can be handled in a similar way. The terms contributing to this class must all involve an $\widetilde{A}_n$
amplitude with the set of legs $\{1,2,\ldots,k\}$ always next to each other, \textit{i.e.} the terms from eq.~\eqref{pureKLT} that
make up this contribution is
\begin{align}
&\sum_{\sigma,\widetilde{\sigma},\alpha}
\widetilde{A}_n(\widehat{n-1},n,\widetilde{\sigma}_{k+1,n-2},\alpha_{2,k},
\widehat{1})
\mathcal{S}[\widetilde{\sigma}_{k+1,n-2}\alpha_{2,k}
|\sigma_{2,n-2}]_{\widehat{p}_1}
A_n(\widehat{1},\sigma_{2,n-2},\widehat{n-1},n)\,,
\end{align}
where we have omitted the overall factor of $(-1)^{n+1}$, which can easily be reinstated into the proof.
We emphasis that since we are considering (A) we have excluded all $\sigma$ permutations that would lead to a $s_{12\ldots k}$ pole
in $A_n$. From this we get the residue
\begin{align}
&\sum_{\sigma,\widetilde{\sigma},\alpha} \frac{\sum_h\widetilde{A}(
\widehat{n-1},n,\widetilde{\sigma}_{k+1,n-2},\widehat{P}^h)
\widetilde{A}(-\widehat{P}^{-h},\alpha_{2,k},\widehat{1})}{s_{12\ldots k}}
\mathcal{S}[\widetilde{\sigma}_{k+1,n-2}\alpha_{2,k}
|\sigma_{2,n-2}]_{\widehat{p}_1} \nonumber \\
&\hspace{10.3cm} \times  A_n(\widehat{1},\sigma_{2,n-2},\widehat{n-1},n)\,,
\label{Ares}
\end{align}
where $\widehat{P} \equiv \widehat{p}_1 + p_2 + \cdots + p_k$, and we have
used the factorization property of color-ordered gauge theory amplitudes in eq.~\eqref{facgluon}.
Also note that the pole $s_{\widehat{1}2\ldots k}$, from the
factorization of $\widetilde{A}_n$, has been replaced with $s_{12\ldots k}$, \textit{i.e.} without the hat on 1,
from the calculation of the residue. Furthermore,
following from the definition of $\mathcal{S}$ in eq.~\eqref{Sn_ft}, we can write
\begin{align}
\mathcal{S}[\widetilde{\sigma}_{k+1,n-2}\alpha_{2,k}
|\sigma_{2,n-2}]_{\widehat{p}_1}
= \mathcal{S}[\alpha_{2,k}|\rho_{2,k}]_{\widehat{p}_1}
\times (\mathrm{a}\,\, \mathrm{factor}\,\,
\mathrm{independent}\,\, \mathrm{of}\, \alpha)\,,
\label{Sfac1}
\end{align}
where $\rho$ denotes the relative ordering of leg
$2,3,\ldots,k$ in the $\sigma$ set. Collecting everything in \eqref{Ares} that involves the $\alpha$ permutation we have
something of the form
\begin{align}
\sum_{\sigma}\sum_h\underbrace{\left( \sum_{\alpha} \widetilde{A}(-\widehat{P}^{-h},
\alpha_{2,k},\widehat{1}) \mathcal{S}[\alpha_{2,k}
|\rho_{2,k}]_{\widehat{p}_1}\right) }_{0}  \times \sum_{\widetilde{\sigma}} [ \cdots ]\,,
\end{align}
where, as we have seen above, the quantity inside $(\cdots)$ vanishes when $\widehat{P}$ is on-shell (to get it in the exact same form as eq.~\eqref{SA} just use the reflection symmetry and eq.~\eqref{STS}.)
We therefore conclude that contributions from (A) vanish altogether. All the non-vanishing stuff must come from (B).
Let us see how this comes about.

Since both $\widetilde{A}_n$ and $A_n$ now contain the pole $s_{12\ldots k}$, they must both have the set of legs $\{1,2,\ldots,k\}$
collected next to each other. The contributing terms, from eq.~\eqref{pureKLT}, then take the form
\begin{align}
&\sum_{\sigma,\widetilde{\sigma},\alpha,\beta}
\widetilde{A}_n(\widehat{n-1},n,\widetilde{\sigma}_{k+1,n-2},
\alpha_{2,k},\widehat{1})
\mathcal{S}[\widetilde{\sigma}_{k+1,n-2}\alpha_{2,k}
|\beta_{2,k}\sigma_{k+1,n-2}]_{\widehat{p}_1} \nonumber \\
&\hspace{9.5cm} \times A_n(\widehat{1},\beta_{2,k},
\sigma_{k+1,n-2},\widehat{n-1},n)\,.
\label{midB}
\end{align}
Like in eq.~\eqref{Sfac1} we will again	 make use of a factorization property of the $\mathcal{S}$ function.
It is not hard to convince oneself that in the limit where $\widehat{P} = \widehat{p}_1 + p_2 + \cdots + p_k$ goes on-shell,
we can write
\begin{align}
\mathcal{S}[\widetilde{\sigma}_{k+1,n-2}\alpha_{2,k}
|\beta_{2,k}\sigma_{k+1,n-2}]_{\widehat{p}_1}
= \mathcal{S}[\alpha_{2,k}|\beta_{2,k}]_{\widehat{p}_1} \times
\mathcal{S}[\widetilde{\sigma}_{k+1,n-2}|\sigma_{k+1,n-2}]_{\widehat{P}}\,.
\end{align}
The residue for $s_{12\ldots k}$ can then be expressed as
\begin{align}
& \frac{1}{s_{12\ldots k}}\sum_h \lim_{z\rightarrow z_{12\ldots k}}\sum_{\alpha,\beta}
\frac{\widetilde{A}(-\widehat{P}^h,\alpha_{2,k},\widehat{1})
\mathcal{S}[\alpha_{2,k}|\beta_{2,k}]_{\widehat{p}_1}
A(\widehat{1},\beta_{2,k},-\widehat{P}^h)}{s_{\widehat{1}2\ldots k}} \nonumber \\
&\times \sum_{\sigma,\widetilde{\sigma}}
\widetilde{A}(\widehat{n-1},n,\widetilde{\sigma}_{k+1,n-2},\widehat{P}^{-h})
\mathcal{S}[\widetilde{\sigma}_{k+1,n-2}|\sigma_{k+1,n-2}]_{\widehat{P}}
A(\widehat{P}^{-h},\sigma_{k+1,n-2},\widehat{n-1},n) \nonumber \\
& + (\mathrm{mixed}\,\,\mathrm{helicity}\,\, \mathrm{terms})\,,
\label{resB}
\end{align}
where the ``mixed helicity terms'' are expressions of the same form but with products between amplitudes with, for instance
$(\widehat{P}^h,\widehat{P}^{-h})$ instead of $(-\widehat{P}^h,-\widehat{P}^{h})$ or $(\widehat{P}^{-h},\widehat{P}^{-h})$
like in line one and two of \eqref{resB}. However, as we discussed in section \ref{vanish_rel}, such terms are identically zero.

What is left is a sum over $\alpha$ and $\beta$, which precisely makes up the
regularized KLT form in eq.~\eqref{newKLT}, \textit{i.e.} it is just
$M_{k+1}(\widehat{1},2,\ldots,k,-\widehat{P}^h)$, and a
sum over $\sigma$ and $\widetilde{\sigma}$ which is an $n-k+1$ point
version of eq.~\eqref{pureKLT} and hence, by induction, equal
to $M_{n-k+1}(k+1,\ldots,\widehat{P}^{-h})$. Altogether \eqref{resB} is
\begin{align}
\sum_h \frac{M_{k+1}(\widehat{1},2,\ldots,k,-\widehat{P}^h)
M_{n-k+1}(k+1,\ldots,\widehat{P}^{-h})}{s_{12\ldots k}}\,.
\end{align}
This is exactly the BCFW-contribution
to the $n$-point gravity amplitude from a $s_{12\ldots k}$ pole, and due to the manifest $(n-3)!$ symmetry
we have also already obtained the contributions for all poles related to these by a permutation of legs
$2,3,\ldots,n-2$.

%===========================================================================
\begin{figure}[t]
\centering
\includegraphics[width=12cm]{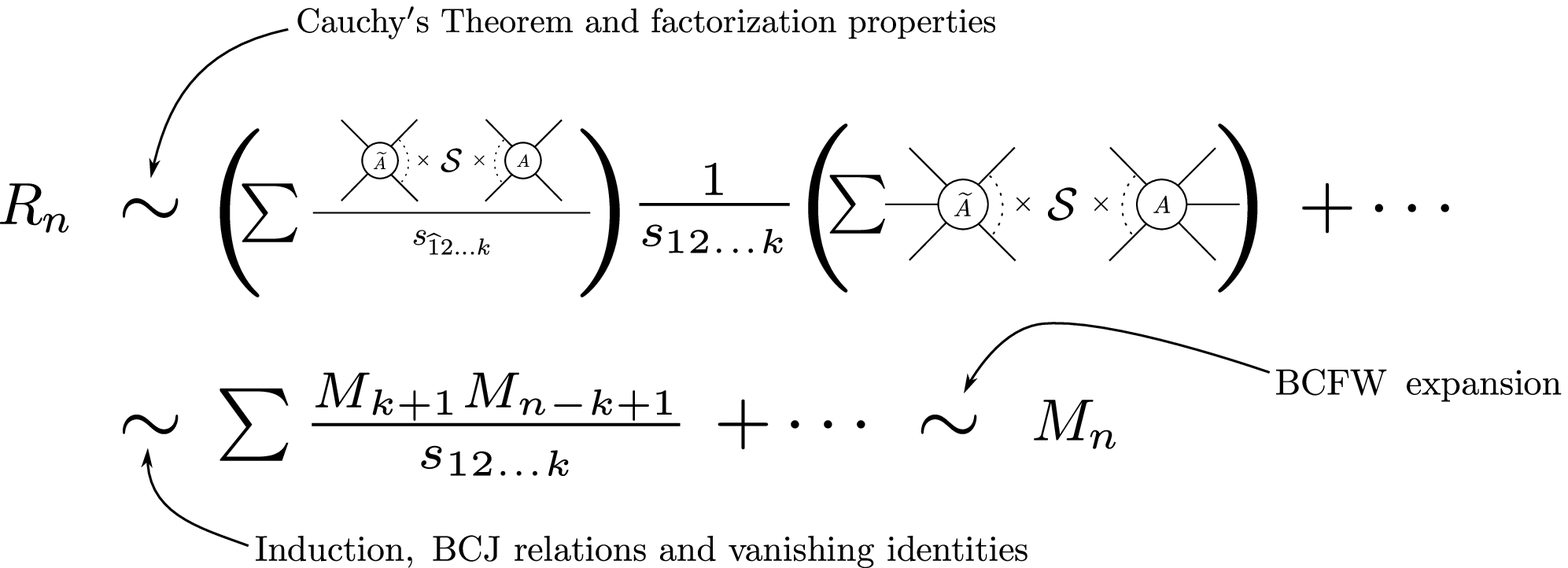}
\caption{\sl Schematic outline of field theory proof.}
\label{fig:field_proof}
\end{figure}
%===========================================================================

We are still not completely done, since the above analysis does not cover the
pole contributions involving both 1 and $n$, \textit{i.e.} poles
of the form $s_{12\ldots k n} = s_{k+1\ldots n-1}$. To
investigate these we use the form in eq.~\eqref{dualKLT}. It is well suited for this
case since here legs 1 and $n$ are always next to each other, and
as we have already mentioned, we are free to use whichever form we want to calculate a residue,
they are just different ways of writing the same quantity. The following calculations
are very similar to what we have already seen so we go through it a bit more briefly.

The (A) part of the residue for pole $s_{12\ldots k n}$ takes the form
\begin{align}
&\sum_{\sigma,\widetilde{\sigma},\alpha}
A_n(\widehat{1},\sigma_{2,n-2},\widehat{n-1},n)
\mathcal{S}[\sigma_{2,n-2}
|\widetilde{\sigma}_{k+1,n-2}\alpha_{2,k}]_{\widehat{p}_{n-1}} \nonumber \\
&\hspace{6cm} \times \sum_h \frac{\widetilde{A}(\widehat{n-1},
\widetilde{\sigma}_{k+1,n-2},\widehat{P}^h)
\widetilde{A}(-\widehat{P}^{-h},\alpha_{2,k},n,\widehat{1})}{s_{12\ldots k n}}\,,
\end{align}
where we again assume the pole appears in the $\widetilde{A}_n$ amplitude.
Using the factorization property
\begin{align}
\mathcal{S}[\sigma_{2,n-2}
|\widetilde{\sigma}_{k+1,n-2}\alpha_{2,k}]_{\widehat{p}_{n-1}}
={}& \mathcal{S}[\rho_{k+1,n-2}
|\widetilde{\sigma}_{k+1,n-2}]_{\widehat{p}_{n-1}}
&\hspace{-0.5cm}\times\, (\mathrm{a}\,\, \mathrm{factor}\,\,
\mathrm{independent}\,\, \mathrm{of}\, \widetilde{\sigma})\,,
\end{align}
where $\rho$ denotes the relative ordering of leg
$k+1,\ldots,n-2$ in the $\sigma$ set, we once again find that these
contributions contain a factor of
\begin{align}
\sum_{\widetilde{\sigma}} \widetilde{A}(\widehat{n-1},
\widetilde{\sigma}_{k+1,n-2},\widehat{P}^h)
\mathcal{S}[\rho_{k+1,n-2}
|\widetilde{\sigma}_{k+1,n-2}]_{\widehat{p}_{n-1}} = 0\,,
\end{align}
that vanishes due to BCJ relations. There is therefore no contribution from (A).

Considering part (B) for the $s_{12\ldots k n}$ pole, the contributing terms from eq.~\eqref{dualKLT} are
\begin{align}
&\sum_{\sigma,\widetilde{\sigma},\alpha,\beta}
A_n(\widehat{1},\beta_{2,k},\sigma_{k+1,n-2},\widehat{n-1},n)
\mathcal{S}[\beta_{2,k}\sigma_{k+1,n-2}
|\widetilde{\sigma}_{k+1,n-2}\alpha_{2,k}]_{\widehat{p}_{n-1}} \nonumber \\
& \hspace{9cm} \times \widetilde{A}_n(\widehat{1},
\widehat{n-1},\widetilde{\sigma}_{k+1,n-2},\alpha_{2,k},n)\,,
\end{align}
with $\mathcal{S}$ satisfying the factorization property (when $\widehat{P}$ goes on-shell)
\begin{align}
\mathcal{S}[\beta_{2,k}\sigma_{k+1,n-2}
|\widetilde{\sigma}_{k+1,n-2}\alpha_{2,k}]_{\widehat{p}_{n-1}}
={}& \mathcal{S}[\sigma_{k+1,n-2}
|\widetilde{\sigma}_{k+1,n-2}]_{\widehat{p}_{n-1}} \times
\mathcal{S}[\beta_{2,k} |\alpha_{2,k}]_{\widehat{P}}\,.
\end{align}
Hence the residue can be written
\begin{align}
&\frac{1}{s_{12\ldots k n}}\sum_h \sum_{\alpha,\beta}
A(\widehat{1},\beta_{2,k},\widehat{P}^h,n)
\mathcal{S}[\beta_{2,k}|\alpha_{2,k}]_{\widehat{P}}
\widetilde{A}(\widehat{1},\widehat{P}^h,\alpha_{2,k},n) \,\,\, \times \nonumber \\
& \lim_{z\rightarrow z_{k+1\ldots n-1}} \sum_{\widetilde{\sigma},\sigma}
\frac{A(-\widehat{P}^{-h},\sigma_{k+1,n-2},\widehat{n-1})
\mathcal{S}[\sigma_{k+1,n-2}
|\widetilde{\sigma}_{k+1,n-2}]_{\widehat{p}_{n-1}}
\widetilde{A}(\widehat{n-1},\widetilde{\sigma}_{k+1,n-2},
-\widehat{P}^{-h})}{s_{k+1\ldots \widehat{n-1}}}\,,
\end{align}
where we have used $s_{\widehat{1}2\ldots k n}=s_{k+1\ldots\widehat{n-1}}$,
and already removed the vanishing 
mixed helicity terms. The first part is
just a lower-point version of eq.~\eqref{dualKLT}, and the
second part the regularized dual KLT form, \textit{i.e.} we get
\begin{align}
\sum_h \frac{M_{k+2}(\widehat{1},2,\ldots,k,n,
\widehat{P}^h)M_{n-k}(k+1,\ldots,\widehat{n-1},
-\widehat{P}^{-h})}{s_{12\ldots k n}}\,.
\end{align}
Once again we obtain the correct BCFW
expansion for all $s_{12\ldots k n}$ poles, and all poles related to
these by a permutation of $2,3,\ldots,n-2$.

We have now covered all residues in eq.~\eqref{cauchyR}, and see that they indeed make up the full
BCFW expansion of the $n$-point gravity amplitude, and therefore $R_n = M_n$, see figure~\ref{fig:field_proof}. Notice how all of the properties/relations
from section~\ref{ftlimit_sec} played an important role; the BCFW method was the main tool for the whole proof, the BCJ relations were needed not only for
showing the equivalence between all $j$-values, but also to argue for the vanishing of contributions from (A), and both the vanishing identities
and regularized KLT relations were important for identifying contributions from (B) with terms from a BCFW expansion of $M_n$. We also stress that
no other crossing symmetry than that which was already manifest has been used, so the identification with a gravity amplitude is also an indirect proof of the total crossing symmetry of the r.h.s. of eq.~\eqref{Mnfinal_ft}.

Maybe the most important thing to take from this proof is, that it very explicitly illustrates how amazingly constrained scattering amplitudes are just from their very general analytical properties.
These constraints are so strong that they basically force perturbative gravity and gauge theories to be related through the KLT relations,
although a priori these theories seem completely unrelated.

\section{The BCJ Representation\label{BCJ_sec}}
In the last couple of years an exciting new structure for gauge theory amplitudes has been discovered \cite{BCJ},
which has also offered a new way of thinking of gravity amplitudes as a ``double-copy'' of gauge theory amplitudes.
One of the more surprising results of this picture is that it seems to generalize to loop level \cite{Bern:2010ue},
whereas the KLT relations by themself only make sense at tree level.
In this section we will briefly review this new and interesting structure.

\subsection{Motivation}
\begin{figure}[t]
\centering
\includegraphics[width=12cm]{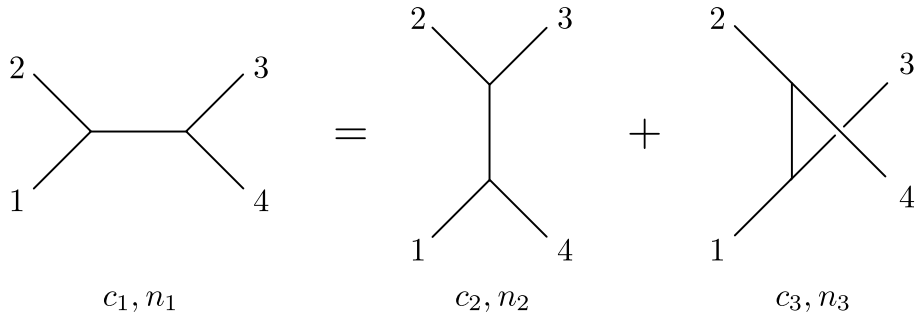}
\caption{\sl Diagrammatic representation of the Jacobi identity. The $c_i$'s can be obtained by dressing the vertices with
structure constants $\tilde{f}^{abc}$, but the diagrams can also be thought of as representing the kinematic numerators $n_i$.}
\label{fig:jacobi}
\end{figure}

The approach can be motivated by making some observations at four points. We start by forcing the color-ordered amplitudes
into a form corresponding to having only anti-symmetric three-point vertices
\begin{align}
A_4(1,2,3,4) = \frac{n_1}{s_{12}} + \frac{n_2}{s_{14}}\,, \quad
A_4(1,2,4,3) = - \frac{n_3}{s_{13}} - \frac{n_1}{s_{12}}\,, \quad
A_4(1,3,2,4) = - \frac{n_2}{s_{14}} + \frac{n_3}{s_{13}}\,.
\label{4ptnumrep}
\end{align}
Any four-point contact terms have simply been absorbed into the numerators, using trivial relations like $s_{12}/s_{12} = 1$,
and the relative signs have been chosen in correspondence
with anti-symmetric cubic vertices. This is quite straightforward to do at four points. In terms of this representation the full color-dressed
gauge theory amplitude takes the form
\begin{align}
A_4^{full} =  \frac{c_1n_1}{s_{12}} + \frac{c_2n_2}{s_{14}} + \frac{c_3n_3}{s_{13}} \,,
\label{4ptfullnumrep}
\end{align}
where the color factors $c_i$ are given by contractions between the structure constants $\tilde{f}^{abc} \equiv i\sqrt{2}f^{abc} =
\mathrm{Tr}([T^a,T^b]T^c)$, with $T^i$ being the generators of the gauge group, as
\begin{align}
c_1 \equiv \tilde{f}^{a_1a_2 b}\tilde{f}^{b a_3 a_4}\,, \quad c_2 \equiv \tilde{f}^{a_2a_3b}\tilde{f}^{ba_4a_1}\,,
\quad c_3 \equiv \tilde{f}^{a_4a_2b}\tilde{f}^{ba_3a_1}\,.
\end{align}
For simplicity we have omitted the coupling constant in eq.~\eqref{4ptfullnumrep}, something we will do
for the remainder of this section.
It is well known that the color factors satisfy the Jacobi identity $c_2+c_3-c_1 = 0$, however, what might be a bit more surprising
is the fact that so do the kinematic numerators, \textit{i.e.} $n_2 + n_3 - n_1 = 0$, see figure~\ref{fig:jacobi}, although this is actually a quite old result \cite{Zhu:1980sz}.
There seems to exist a kind of \textit{color-kinematic duality} in the structure of four-point amplitudes.

In addition, if we use the representation of amplitudes in eq.~\eqref{4ptnumrep}, plug it into the four point KLT relation in eq.~\eqref{4ptKLT}
and use the Jacobi identity for the numerators, we get
\begin{align}
M_4 =  \frac{n_1\widetilde{n}_1}{s_{12}} + \frac{n_2\widetilde{n}_2}{s_{14}} + \frac{n_3\widetilde{n}_3}{s_{13}} \,,
\label{4ptMnn}
\end{align}
where $\widetilde{n}_i$ is just the numerators belonging to $\widetilde{A}_4$ written in a similar representation.
Compared to eq.~\eqref{4ptfullnumrep} we have basically just replaced the color factors $c_i$ with kinematic numerators $\widetilde{n}_i$.
This nicely captures the statement of gravity being a ``double-copy'' of gauge theory.

%===========================================================================
%===========================================================================

\subsection{At General $n$-point Tree and Loop Level}
One could think that the above duality between color factors and kinematic numerators is merely a curious coincidence at four points, however,
Bern, Carrasco and Johansson promoted it to a general principle saying; it is \textit{always} possible to represent an $n$-point
gauge theory amplitude as a sum over all distinct $n$-point diagrams with only anti-symmetric cubic vertices, and in such a way that the
kinematic numerators $n_i$ satisfy the same Jacobi identities
as the color factors $c_i$. In detail, this is to write the full tree-level amplitude in the form
\begin{align}
A_n^{full}(1,2,\ldots,n) =  \sum_{i} \frac{c_in_i}{(\prod_j s_j)_i} \,,
\label{Acn}
\end{align}
where the color-kinematic duality
\begin{align}
c_{\alpha}+c_{\beta}-c_{\gamma} = 0 \qquad \longrightarrow \qquad  n_{\alpha}+n_{\beta}-n_{\gamma} = 0\,,
\label{colkindual}
\end{align}
is satisfied. The sum in eq.~\eqref{Acn} is over all different diagrams, only containing cubic vertices,
and $(\prod_j s_j)_i$ is the corresponding pole structure of diagram $i$. It should be noted, that in the four point case
the duality in \eqref{colkindual} was automatically satisfied once the amplitude was forced into the form of eq.~\eqref{4ptfullnumrep},
this is \textit{not} the case for general $n$-point amplitudes. To write an amplitude in the form of eq.~\eqref{Acn} is not difficult, but
to \textit{also} have the numerators satisfy \eqref{colkindual} is a highly non-trivial task \cite{BCJstructure,Mafra:2011kj}.
Recently there has been some progress in understanding the duality at a deeper level, looking for an underlying kinematic group \cite{Bern:2011ia}.

It is also worth mentioning that the BCJ relations reviewed in section~\ref{BCJ_relations}, although having a life of its own now,
was originally found as a consequence of the color-kinematic duality. The constraint of a Jacobi structure on the kinematic numerators lead to
those relations among the color-ordered amplitudes.

Similarly to eq.~\eqref{4ptMnn}, once a representation satisfying the color-kinematic duality is obtained, the $n$-point gravity amplitude
is given by
\begin{align}
M_n(1,2,\ldots,n) = \sum_{i} \frac{\widetilde{n}_in_i}{(\prod_j s_j)_i} \,,
\label{nptMnn}
\end{align}
we just replace the color factors $c_i$ in eq.~\eqref{Acn} with kinematic numerators $\widetilde{n}_i$ from a gauge theory amplitude
written in a similar representation  \cite{BCJ,Bern:2010yg}.

Remarkably, there are strong indications that the color-kinematic duality, and the connection to gravity through ``squaring'' of numerators,
has an extension to loop level \cite{Bern:2010ue}. The statement is that the $n$-point gauge theory amplitude at $L$ loops can be written in the form
\begin{align}
(-i)^L A_n^{loop} = \sum_{i}\int \prod_{l=1}^L \frac{d^Dp_l}{(2\pi)^D} \frac{1}{S_i} \frac{c_in_i}{(\prod_j s_j)_i} \,,
\end{align}
where the sum is over all $n$-point $L$-loop diagrams with cubic vertices, $S_i$ is the symmetry factor of diagram $i$, and
the numerators $n_i$ again satisfy the color-kinematic duality in \eqref{colkindual}. Note that the
numerators are something belonging to the integrand. The conjecture for the $n$-point $L$-loop gravity amplitude is then
\begin{align}
(-i)^{L+1} M_n^{loop} = \sum_{i}\int \prod_{l=1}^L \frac{d^Dp_l}{(2\pi)^D} \frac{1}{S_i} \frac{\widetilde{n}_in_i}{(\prod_j s_j)_i} \,,
\end{align}
with the same sum as in the gauge theory case. 
This is not only a theoretically interesting extension of the connection between perturbative gravity and gauge theory
beyond tree level, but
together with generalized unitarity cut methods \cite{Bern:1994zx} can have significant
implications for multi-loop calculations, see \textit{e.g.} \cite{Bern:2010tq}, which
makes it a most interesting structure.

\section{Conclusions\label{conclusions}}
In this review we have taken a close look at the KLT relations. The first part involved a rather detailed derivation from string theory,
showing how to factorize the \textit{closed} string integrals into two separate sectors which were then deformed into expressions corresponding
to products of \textit{open} string amplitudes. One of the important steps was to keep track of phase factors.
These depended a lot on the way we chose to close the integral contours around the
branch cuts. However, we saw that this freedom could be nicely incorporated into a kinematic $\mathcal{S}$ function leading to very
compact $n$-point expressions.

The second part focused on the KLT relations in the field theory limit. We investigated some of the amazing properties
these relations exhibit and introduced tools that would be needed for a purely field theoretical proof. In particular, we saw how
the BCJ relations connected all the different expressions obtained in string theory from closing contours in different ways. In light of
the monodromy relations, the connection between deforming integral contours in string theory and the BCJ relations in field theory
is now well understood. Indeed, the BCJ and KLT relations are much more closely related than one would first have anticipated.
We ended the second part of the review with a proof of the $n$-point KLT relations from a purely field theoretical point of view.
Like most other properties presented it was based on very general analytical structures of scattering amplitudes.

In the third and final part, we reviewed a very recently discovered way of representing gauge theory amplitudes, following from a
color-kinematic duality, and we saw how this implied a new representation for gravity amplitudes as well. Probably one of the
most exciting things about this approach is its generalization to all loop-order amplitudes, rendering a new way
of doing loop calculations. Only the future will show what additional secrets the connection between gravity and gauge theory has in store
for us.

\section*{Acknowledgments}
We thank Zvi Bern, Emil Bjerrum-Bohr, Poul Henrik Damgaard, Bo Feng, Pierre Vanhove and Cristian Vergu for discussions and comments.
Special thanks go to Kemal Ozeren for his comments on an early draft. This research was supported in part by the National Science Foundation under Grant No. NSF PHY05-51164.

\appendix

\section{The Complex Power Function and Branch Cuts\label{branchcuts}}
The complex power function $z^c$ is in general a \textit{multi-valued} function. As a simple example consider
$z^{1/2}$ and take $z= |z|e^{i\theta} = |z|e^{i\theta + 2\pi i}$, then
\begin{align}
z^{1/2} ={}& \left( |z| e^{i\theta} \right)^{1/2} = |z|^{1/2} e^{i\theta/2}\,, \nonumber \\
z^{1/2} ={}& \left( |z| e^{i\theta + 2\pi i} \right)^{1/2} =  |z|^{1/2} e^{i\theta/2} e^{i\pi} = -|z|^{1/2} e^{i\theta/2}\,,
\end{align}
which is obviously two different results for the same $z$. 

In order to have a well defined function, here meaning \textit{single-valued}, we need to impose a \textit{branch cut}
on the complex plane. Most often, one choose the branch cut to lie on the negative real axis, that is we restrict the
power function to
\begin{align}
z^c= |z|^c e^{ci\theta}  \,, \qquad -\pi < \theta \leq \pi \,.
\label{power_cut}
\end{align}
What this means is; (1) take any complex number $z$, (2) write it up in polar coordinates,
\textit{i.e.} $z=|z|e^{i\theta}$, with $\theta \in\, ]-\pi,\pi]$, (3) then the function gives you back the complex number $|z|^c e^{ci\theta}$.

It might seem like we are being overly cautions, but it is so easy to just start using rules we know from ordinary
power functions, like $(z_1z_2)^c = z_1^cz_2^c$, which is \textit{not} always true in the complex case.

\subsection{Phase Factors}
Let us see how the above branch cut dictates the phase factors in eq.~\eqref{phase_factors}. That is, 
we are considering eq.~\eqref{power_cut} for some $z_0$ with $\mathrm{Re}(z_0)<0$. If $\mathrm{Im}(z_0) \geq 0$ we have
\begin{align}
(z_0)^c = |z_0|^c e^{ci\theta_0}  \,, \qquad \pi/2 < \theta_0 \leq \pi \,.
\end{align}
If we now look at the power of $-z_0$ we get
\begin{align}
(-z_0)^c = |z_0|^c e^{ci(\theta_0-\pi)} = e^{-i\pi c}|z_0|^c e^{ci\theta_0} = e^{-i\pi c} (z_0)^c  \,.
\end{align}
Note that we \textit{have to} take the argument for $(-z_0)$ as $\theta_0 -\pi$ in order not to fall outside the $]-\pi,\pi]$ range.
We can then conclude that
\begin{align}
(z_0)^c =  e^{i\pi c} (-z_0)^c\,.
\end{align}
In particular, we see that $e^{-i\pi c} (-z_0)^c  =  e^{-2i\pi c} (z_0)^c \neq (z_0)^c$ in general, and
we therefore have to pick the phase $e^{i\pi c}$, like in eq.~\eqref{phase_factors}.

Likewise, if $\mathrm{Im}(z_0) < 0$ we have
\begin{align}
(z_0)^c = |z_0|^c e^{ci\theta_0}  \,, \qquad -\pi < \theta_0 < -\pi/2 \,,
\end{align}
and for $-z_0$
\begin{align}
(-z_0)^c = |z_0|^c e^{ci(\theta_0+\pi)} = e^{i\pi c}|z_0|^c e^{ci\theta_0} = e^{i\pi c} (z_0)^c  \,,
\end{align}
hence
\begin{align}
(z_0)^c = e^{-i\pi c} (-z_0)^c\,.
\end{align}
Again we see that $(z_0)^c \neq e^{i\pi c} (-z_0)^c$, so our branch cut implies the phases used in eq.~\eqref{phase_factors}.

The careful reader might also have noticed that in eq.~\eqref{split_one} and \eqref{split_two} we actually did use $(z_1z_2)^c = z_1^cz_2^c$,
so how do we know this is allowed here without any modifications in terms of phase factors? The reason is, that in this case $z_1$ and $z_2$ have
opposite sign on the imaginary part. For instance, say $\mathrm{Im}(z_1) < 0$ and $\mathrm{Im}(z_2) > 0$, then $\theta_1 \in ]-\pi, 0[$ and
$\theta_2 \in ]0,\pi[$, which imply $(\theta_1 + \theta_2) \in\, ]-\pi,\pi[$, and therefore in this situation
$ z_1^c z_2^c$ is equal to $(z_1z_2)^c$.

\end{document}